\documentclass[journal]{IEEEtran}
\usepackage{amsmath, amssymb, amsthm}
\usepackage{algorithm}
\usepackage{algpseudocode}
\usepackage{graphicx}
\usepackage{subcaption}
\usepackage{tabularx}
\usepackage{amssymb} 
\usepackage{setspace}
\usepackage{multicol}
\usepackage[font=small,labelfont=bf]{caption}
\usepackage{listings}
\usepackage{verbatim}
\usepackage{fancyvrb}
\usepackage{xcolor}
\usepackage{booktabs}
\usepackage{graphicx}
\usepackage{pifont}
\usepackage{algpseudocode}
\usepackage[normalem]{ulem}
\usepackage[utf8]{inputenc}
\DeclareUnicodeCharacter{221E}{\ensuremath{\infty}}
\usepackage[hidelinks]{hyperref}
\hypersetup{
    colorlinks=true,
    linkcolor=blue,
    citecolor=blue,
    urlcolor=blue
}
\newcommand{\cmark}{\ding{51}}
\newcommand{\xmark}{\ding{55}}

\begin{document}

\title{Hallucination Cascade: Analyzing Error Propagation in Multi-Agent LLM Systems}

\newcommand{\Arghavan}[1]{\textcolor{brown}{{\it [Arghavan: #1]}}}

\author{
Saeid~Jamshidi,
Arghavan~Moradi~Dakhel,
Kawser~Wazed~Nafi and
Foutse~Khomh

\thanks{
S. Jamshidi, A. Moradi Dakhel, K. W. Nafi, and F. Khomh are with the SWAT Laboratory,
Polytechnique Montréal, Montréal, QC, Canada
(e-mail: \{saeid.jamshidi, arghavan.moradi-dakhel, kawser.wazed-nafi, foutse.khomh\}@polymtl.ca).
}%

}

\maketitle

\begin{abstract}
Large Language Models (LLMs) have demonstrated capabilities in natural language generation, but they remain vulnerable to hallucinations, producing fluent responses that contain unsupported and factually incorrect claims. Most existing studies evaluate hallucination as a static property of isolated model outputs. However, in multi-agent LLM systems, generated responses are repeatedly exchanged between agents, refined across stages, and reused as context for subsequent reasoning. In this setting, hallucination becomes a dynamic process shaped by interaction history, cascade depth, and model heterogeneity. This paper analyzes hallucination dynamics in multi-agent LLM cascades by tracking claim-level factual inconsistency across sequential agent interactions. We conduct 500 cascade experiments across 10 knowledge domains using three heterogeneous models: GPT-5.3, DeepSeek-V3, and LLaMA-3-70B-Instruct, yielding 1,250 evaluated responses. The results show that deeper cascades reduce the normalized hallucination score from 0.422 at the first agent to 0.272 at the final agent in 3-agent chains, with an amplification factor of 0.644, indicating net attenuation, as values below one represent reduced hallucination across the cascade. However, this reduction is accompanied by a decline in factual accuracy from 0.789 to 0.769, revealing a trade-off between hallucination suppression and factual preservation. Transition-level analysis shows that each agent-to-agent refinement reduces hallucination by an average of $0.072$, while introducing small but consistent losses in factual consistency and response quality. Model-level results reveal distinct reliability-efficiency trade-offs: LLaMA-3-70B-Instruct achieves the lowest hallucination score, whereas GPT-5.3 provides faster generation but exhibits a higher hallucination rate. Domain-level analysis further shows that hallucination is sensitive to topic complexity, with lower scores in well-grounded scientific domains and higher scores in more abstract knowledge domains.
\end{abstract}


\section{Introduction}
\label{Introduction}
Large Language Models (LLMs) have substantially advanced Natural Language Generation (NLG), enabling fluent and contextually coherent text generation across applications such as question answering, dialogue systems, and decision support \cite{karanikolas2023large,ren2024advancements,hadi2023large}. As LLMs are increasingly deployed in real-world settings, factual correctness and reliability have become as important as linguistic quality \cite{gu2023don,augenstein2024factuality,mcintosh2025inadequacies}. Despite their generative capabilities, LLMs remain prone to hallucination \cite{lu2026loki}, producing plausible but factually incorrect content that lacks support from verifiable evidence \cite{gao2025h,kulkarni2025scientific}. This limitation raises major concerns for trust, safety, and robustness, particularly in high-stakes domains such as healthcare and engineering systems~\cite{survey_hallu,medical_hallu}.
Prior work has made substantial progress in defining, evaluating, detecting, and mitigating hallucinations \cite{huang2025survey,tonmoy2024comprehensive}. Existing studies distinguish between intrinsic hallucinations, which contradict the input, and extrinsic hallucinations, which are unsupported by external evidence~\cite{survey_hallu}. Benchmarks such as TruthfulQA show that larger models can still generate convincing but incorrect answers, indicating that scaling alone does not guarantee factual reliability~\cite{truthfulqa}. Model-based evaluators such as G-Eval assess factuality and response quality through structured semantic judgment~\cite{geval}, while detection methods such as SelfCheckGPT identify hallucinated content by measuring inconsistencies across multiple sampled responses~\cite{selfcheckgpt}. Grounding-based methods further reduce hallucinations by incorporating external knowledge sources, such as knowledge graphs and structured semantic resources~\cite{kg_llm,kg_sysml}. However, most existing approaches treat hallucination as a static property of individual outputs. They typically evaluate whether a single response is factually correct after generation, without modeling how factual errors evolve across multi-step reasoning and multi-agent interaction. This assumption becomes restrictive in emerging multi-agent LLM systems, where one agent's output is reused as context by subsequent agents. In such settings, an early hallucinated claim may be preserved, softened, corrected, amplified, transformed, and replaced as it moves through the cascade. A static evaluator can score the final response, but it cannot determine where the error emerged, how it changed across agents, and whether later refinement improved and degraded factual reliability. This paper addresses this gap by modeling hallucination as a dynamic stochastic process in multi-agent LLM cascades. We decompose generated responses into atomic claims, evaluate each claim using rule-based grounding and LLM-based semantic validation, and track claim-level factual inconsistency across sequential agents. This enables hallucination to be analyzed as a trajectory rather than a single output-level score. We further define propagation metrics to quantify attenuation, amplification, recovery, semantic drift, stability, and failure behavior across cascade stages. By shifting the analysis from isolated responses to inter-agent propagation, this work provides a systematic method for studying how hallucinations emerge, evolve, and impact reliability in multi-agent LLM systems.

The main contributions of this work are summarized as follows:
\begin{itemize}
    \item \textbf{Stochastic Modeling of Hallucination Dynamics:} We model hallucination as a stochastic process that evolves over sequential interactions, enabling the characterization of temporal dependencies, inter-step error propagation, attenuation, amplification, recovery, and stability in multi-agent LLM systems.
    
    \item \textbf{Claim-Level Hybrid Estimation of Hallucination:} We introduce a fine-grained estimation approach that decomposes generated outputs into atomic claims and quantifies hallucination by integrating rule-based grounding with LLM-based semantic validation, clarifying factual support and semantic plausibility at the claim level.
    
    \item \textbf{Quantitative Analysis of Multi-Agent Error Propagation:} We construct multi-agent cascades and define propagation metrics to measure how hallucinations evolve across agents, capturing system-level behavior through cross-model cascades, long-chain analysis, injected hallucination experiments, error recovery analysis, and failure case analysis.
\end{itemize}

The remainder of this paper is organized as follows. Section~\ref{Related Work} reviews existing hallucination studies in LLMs. Section~\ref{Methodology} presents the proposed dynamic formulation for multi-agent hallucination analysis. Section~\ref{sec:experimental_setup} describes the experimental design and evaluation metrics. Section~\ref{Experimental Results and Analysis} reports hallucination propagation results and model-level trade-offs. Section~\ref{Discussion} discusses the main findings, followed by limitations in Section~\ref{Limitations} and future work in Section~\ref{Future Work}. Section~\ref{Conclusion} concludes the paper.

\section{Related Work}
\label{Related Work}

This section reviews prior work on hallucination in LLMs, covering definitions, evaluation, detection, grounding-based mitigation, and domain-specific reliability challenges.

\subsection{Hallucination in LLMs}
LLMs can generate fluent and coherent text, but they remain vulnerable to hallucination, producing factually incorrect and unsupported content. Prior work distinguishes between intrinsic hallucinations, which contradict the input, and extrinsic hallucinations, which lack support from verifiable knowledge~\cite{survey_hallu}. Hallucination has also been described as a plausible but erroneous output that can be difficult to distinguish from correct information~\cite{medical_hallu}. This is especially problematic because hallucinated responses often appear linguistically confident, increasing the risk of user trust in incorrect content.

\subsection{Benchmarking and Evaluation of Hallucination}
Hallucination evaluation is challenging because factual correctness depends on semantic grounding rather than surface similarity. TruthfulQA~\cite{truthfulqa} evaluates whether models reproduce common misconceptions and shows that larger models may still generate convincing but incorrect answers. Model-based evaluators such as G-Eval~\cite{geval} use LLMs to assess output quality through structured scoring, but they raise concerns about evaluator bias, self-consistency, and shared failure patterns between the judge and evaluated model.

\subsection{Hallucination Detection Methods}
Detection methods aim to identify hallucinated content after generation. SelfCheckGPT~\cite{selfcheckgpt} detects hallucination by measuring inconsistency across multiple sampled responses, based on the idea that factual knowledge produces stable outputs while hallucinated content leads to divergence. However, most detection methods operate at the response and sentence level and do not explicitly model claim-level structure, propagation, and temporal evolution across reasoning steps.

\subsection{Knowledge Grounding and Mitigation Strategies}
Grounding-based methods reduce hallucination by incorporating external knowledge sources. Knowledge graph approaches~\cite{kg_llm} improve factual consistency by providing structured relationships among entities, while hierarchical knowledge-based reasoning methods~\cite{kg_sysml} guide generation with structured prompts and semantic constraints. Although these methods improve reliability, they are usually applied at a single generation stage and do not explain how residual hallucinated claims evolve across subsequent agents.

\subsection{Domain-Specific Hallucination Challenges}
Hallucination is especially concerning in high-stakes domains such as healthcare and engineering. In medical AI, plausible but incorrect outputs may directly impact clinical decision-making~\cite{medical_hallu}. In engineering tasks, hallucinations can produce semantically invalid outputs even when the response appears syntactically correct~\cite{kg_sysml}. These challenges show the need to study hallucination beyond isolated responses, particularly in systems where intermediate outputs are reused across agents, tools, and reasoning stages.\\

The literature synthesis demonstrates that, although prior work has made significant progress in defining, evaluating, detecting, and mitigating hallucinations, most approaches treat hallucination as a static property of individual outputs. Existing methods largely focus on single-step generation and do not capture how hallucinated information evolves during multi-step reasoning and agent interaction. Moreover, current evaluation and mitigation strategies lack mechanisms for modeling temporal error propagation, recovery, attenuation, amplification, and long-chain stability. To address these limitations, this work models hallucination as a dynamic stochastic process and analyzes its propagation within multi-agent LLM systems. By capturing hallucination trajectories and interaction impacts, our approach enables a deeper understanding of error amplification, attenuation, recovery behavior, failure dynamics, and system-level robustness.

\section{Methodology}
\label{Methodology}
\begin{figure*}[t]
\centering
\includegraphics[width=0.70\textwidth]{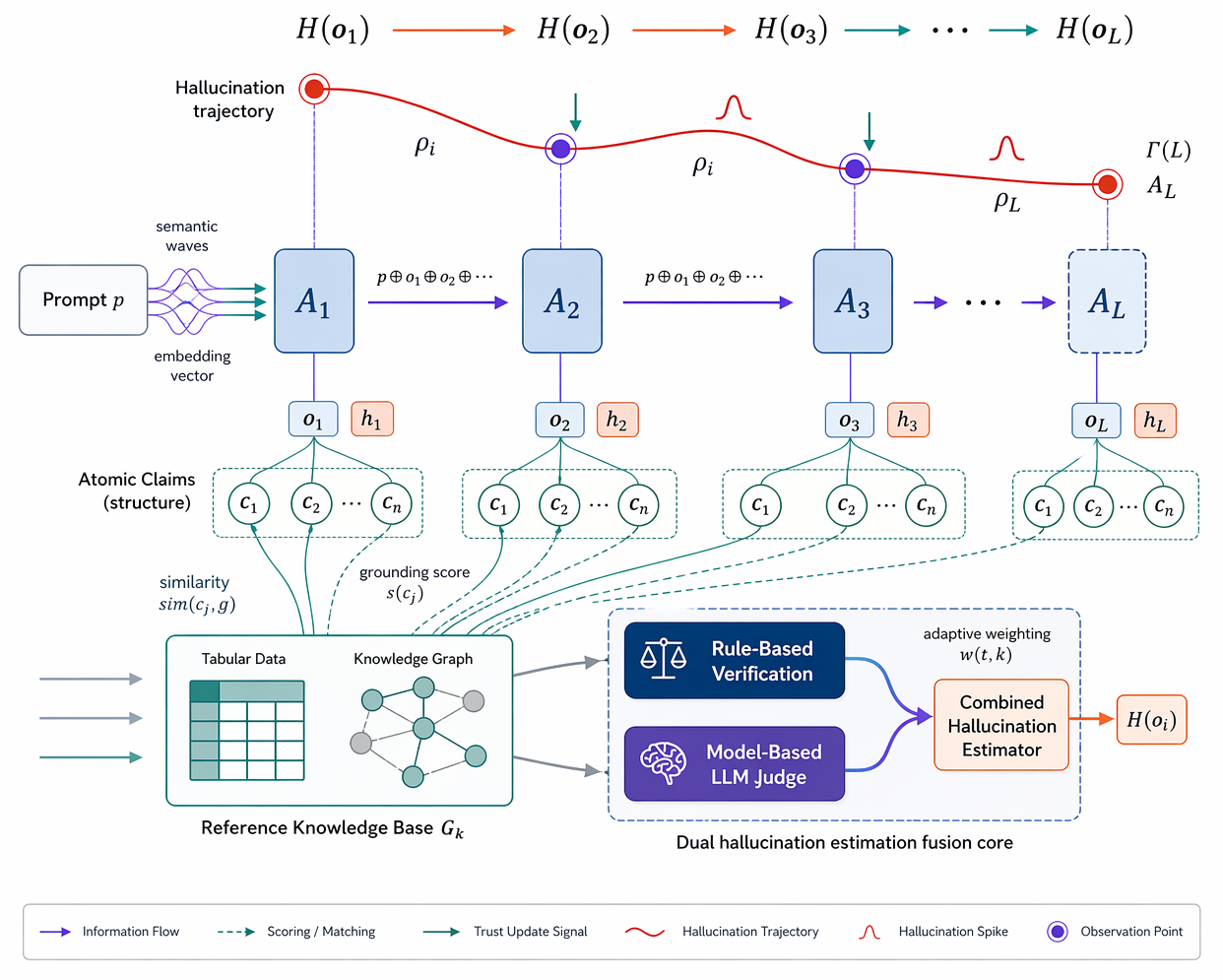}
\caption{Overview of the multi-agent cascade for claim-level hallucination estimation and propagation.}
\label{fig:cascade_pipeline}
\end{figure*}

This section presents a formal, operational method for quantifying hallucinations and tracking their propagation in multi-agent LLM systems. Unlike conventional evaluations that assess isolated outputs, we treat hallucination as a dynamic process that evolves across sequential agent interactions. The method is model-agnostic and can be applied to heterogeneous LLMs, including GPT-5.3, LLaMA, and DeepSeek-V3.
As shown in Figure~\ref{fig:cascade_pipeline}, each generated response is decomposed into verifiable atomic claims. Each claim is evaluated using rule-based grounding and LLM-based semantic validation. Claim-level scores are then aggregated into response-level hallucination scores and tracked across agents, enabling analysis of attenuation, amplification, recovery, semantic drift, long-chain stability, and failure cases.

\subsection{Claim Extraction and Representation}
\label{sec:claim_extraction}
A key step is decomposing each generated response into atomic claims. Given a response $t \in \mathcal{T}$, the claim set is defined as:
\begin{equation}
\mathcal{C}(t) = \{c_1, c_2, \ldots, c_n\}
\end{equation}
where each claim $c_j$ is a minimal, self-contained, and verifiable factual statement.
Claim extraction combines LLM-based semantic segmentation with deterministic refinement:
\begin{equation}
\mathcal{C}(t) = f_{\text{rule}}\big(f_{\text{LLM}}(t)\big)
\end{equation}
where $f_{\text{LLM}}(\cdot)$ identifies candidate factual propositions and $f_{\text{rule}}(\cdot)$ normalizes them into consistent atomic claims. The refinement step applies sentence boundary detection, duplicate removal, clause completion, coreference normalization, and filtering of non-factual fragments. The complete extraction operator is:
\begin{equation}
\mathcal{C}(t) = f_{\text{extract}}(t).
\end{equation}
For example, the response:
\begin{quote}
``Photosynthesis occurs in chloroplasts and produces glucose using sunlight.''
\end{quote}
is separated into:
\begin{equation}
\begin{aligned}
\{ &\text{Photosynthesis occurs in chloroplasts}, \\
  &\text{Photosynthesis produces glucose using sunlight} \}
\end{aligned}
\end{equation}
This decomposition enables fine-grained factuality assessment and supports tracking how claims are preserved, corrected, weakened, amplified, and semantically transformed across cascade stages.

\subsection{Overview of the Approach}
\label{sec:approach_overview}
The proposed method consists of three connected stages:
\begin{enumerate}
    \item \textbf{Claim Decomposition:} Each response is converted into independently verifiable atomic claims, reducing sentence-level and paragraph-level error masking.

    \item \textbf{Claim-Level Grounding:} Each claim is compared with verified domain-specific reference facts using semantic similarity and hallucination-likelihood estimation.

    \item \textbf{Aggregation and Propagation:} Claim-level scores are aggregated into response-level hallucination scores and tracked across agents to capture attenuation, amplification, recovery, semantic drift, and failure behavior.
\end{enumerate}
Furthermore, claim decomposition, grounding, probabilistic scoring, and sequential modeling allow hallucination to be analyzed as a stochastic process across agent interactions rather than as a static property of isolated outputs. This supports the evaluation of cross-model cascades, long-chain behavior, injected-hallucination propagation, recovery patterns, and failure cases.

\subsection{Formal Problem Definition}
\label{sec:formal_problem}
Let $\mathcal{T}$ denote the space of generated texts and $\mathcal{K}$ the set of knowledge domains. For a generated response $t \in \mathcal{T}$, the semantic content is decomposed into atomic claims:
\begin{equation}
\mathcal{C}(t) = \{c_1, c_2, \ldots, c_n\}
\end{equation}
where each claim $c_j$ is a minimal, self-contained, and verifiable factual statement. This decomposition follows the hybrid extraction process in Section~\ref{sec:claim_extraction}, in which an instruction-tuned LLM identifies candidate factual propositions, and deterministic post-processing removes redundancy, resolves co-reference, and enforces structural consistency.
For each domain $k \in \mathcal{K}$, let $\mathcal{G}_k=\{g_1,\ldots,g_m\}$ denote verified reference facts used for grounding. Each claim $c_j$ is matched to its most relevant reference fact:
\begin{equation}
\pi(c_j)=\arg\max_{g \in \mathcal{G}_k}\mathrm{sim}(c_j,g)
\end{equation}
where semantic similarity is computed using cosine similarity between claim and reference embeddings:
\begin{equation}
\mathrm{sim}(c_j,g)=
\frac{\mathbf{e}_{c_j}\cdot \mathbf{e}_g}
{\|\mathbf{e}_{c_j}\|\|\mathbf{e}_g\|}.
\end{equation}
To account for grounding uncertainty, each claim is mapped to a probability distribution over reference facts:
\begin{equation}
P_{c_j}(g)=
\frac{\exp(\mathrm{sim}(c_j,g))}
{\sum_{g' \in \mathcal{G}_k}\exp(\mathrm{sim}(c_j,g'))}.
\end{equation}
The reference distribution is defined as a one-hot distribution centered on the selected grounding fact:
\begin{equation}
P_{\pi(c_j)}(g)=
\begin{cases}
1 & \text{if } g=\pi(c_j), \\
0 & \text{otherwise}.
\end{cases}
\end{equation}
The claim-level hallucination score is:
\begin{equation}
h(c_j)=
\begin{cases}
\delta_{JS}(P_{c_j}\parallel P_{\pi(c_j)})
& \text{if } \max_g \mathrm{sim}(c_j,g)\geq \tau, \\
1 & \text{otherwise},
\end{cases}
\end{equation}
where $\tau \in [0,1]$ is an empirically calibrated grounding threshold. Claims below this threshold are treated as unsupported, while grounded claims are scored using Jensen--Shannon divergence:
\begin{equation}
\begin{aligned}
\delta_{JS}(P \parallel Q) &=
\frac{1}{2}D_{KL}(P \parallel M)
+
\frac{1}{2}D_{KL}(Q \parallel M), \\
M &= \frac{1}{2}(P+Q).
\end{aligned}
\end{equation}
The response-level hallucination score is the mean claim-level score:
\begin{equation}
\mathcal{H}(t,k)=
\frac{1}{|\mathcal{C}(t)|}
\sum_{c_j \in \mathcal{C}(t)} h(c_j),
\end{equation}
which provides a normalized measure of factual inconsistency for response $t$ in domain $k$. For a model $M$ and prompt $p$, the expected hallucination is:
\begin{equation}
\mathcal{H}(M(p),k)=
\mathbb{E}_{t \sim P_M(\cdot \mid p)}
[\mathcal{H}(t,k)].
\end{equation}
This definition supports the construction and analysis of hallucination trajectories across multi-agent cascades.

\subsection{Cascade System}
\label{sec:cascade_system}
We model the multi-agent LLM system as a sequential cascade in which information and factual errors may propagate across agents. Each agent $A_i$ receives the original prompt and the accumulated outputs from previous agents, then generates an updated response. This allows hallucination to be analyzed as a time-dependent signal rather than as an isolated output-level score.
A cascade of length $L$ is defined as:
\begin{equation}
\mathcal{C} = (A_1, A_2, \ldots, A_L)
\end{equation}
where each agent is an independent generative model involved in multi-step reasoning. At step $i$, the agent's output is generated as:
\begin{equation}
o_i \sim P_{M_i}(\cdot \mid p, o_1, \ldots, o_{i-1})
\end{equation}
and equivalently:
\begin{equation}
o_i = M_i\left(p \oplus \left(\bigoplus_{j=1}^{i-1} o_j\right)\right)
\end{equation}
where $\oplus$ denotes concatenation. This accumulated context enables refinement but also creates a pathway for earlier factual errors to impact later generations.
The hallucination trajectory across the cascade is defined as:
\begin{equation}
\mathbf{h} =
\left(\mathcal{H}(o_1), \mathcal{H}(o_2), \ldots, \mathcal{H}(o_L)\right)
\end{equation}
where each element represents the hallucination score of an agent output. The transition between consecutive stages is modeled as:
\begin{equation}
\mathcal{H}(o_i) =
f_i\left(\mathcal{H}(o_{i-1}), \mathcal{H}(p)\right) + \epsilon_i
\end{equation}
where $\epsilon_i \sim \mathcal{N}(0,\sigma_i^2)$ captures model-induced variability.
Propagation is measured using local and global metrics. The relative propagation between consecutive agents is:
\begin{equation}
\rho_i =
\frac{\mathcal{H}(o_i) - \mathcal{H}(o_{i-1})}
{\mathcal{H}(o_{i-1}) + \varepsilon}
\end{equation}
where $\varepsilon=10^{-8}$ ensures numerical stability. Positive values indicate amplification, while negative values indicate attenuation. The cumulative change is:
\begin{equation}
\Gamma(L) = \mathcal{H}(o_L) - \mathcal{H}(o_1)
\end{equation}
and the global amplification factor is:
\begin{equation}
\mathcal{A}_L =
\begin{cases}
\frac{\mathcal{H}(o_L)}{\mathcal{H}(o_1)} & \text{if } \mathcal{H}(o_1) > 0, \\
\mathcal{H}(o_L) & \text{otherwise}.
\end{cases}
\end{equation}
Values of $\mathcal{A}_L<1$ indicate net attenuation, values near one indicate preservation, and values above one indicate amplification. Moreover, $\mathbf{h}$, $\rho_i$, $\Gamma(L)$, and $\mathcal{A}_L$ provide the main metrics for analyzing hallucination propagation across multi-agent cascades.
Moreover, the final hallucination score combines rule-based grounding and LLM-based semantic assessment through adaptive fusion:
\begin{equation}
\mathcal{H}(t,k) =
w(t,k)\cdot \mathcal{H}_{\text{rule}}(t,k)
+
\left(1-w(t,k)\right)\cdot \mathcal{H}_{\text{LLM}}(t,k)
\end{equation}
where $w(t,k) \in [0,1]$ controls the contribution of each component. Higher values favor rule-based grounding when reference support is reliable, while lower values favor the LLM-based judge when grounding evidence is weak, sparse, and incomplete.
The adaptive weight is defined as:
\begin{equation}
w(t,k) =
\frac{1}{1+\exp(-\gamma \cdot \kappa(t,k))}
\end{equation}
where $\gamma>0$ controls transition sensitivity, and $\kappa(t,k)$ measures grounding reliability:
\begin{equation}
\kappa(t,k) =
\frac{1}{|\mathcal{C}(t)|}
\sum_{c_j \in \mathcal{C}(t)} s(c_j)
-
\lambda \cdot \mathrm{Var}(s(c_1), \ldots, s(c_n)).
\end{equation}
The first term captures average semantic support, while the variance term penalizes unstable grounding across claims. Thus, high $\kappa(t,k)$ indicates consistent grounding, whereas low $\kappa(t,k)$ indicates weak and fragmented support.
The sigmoid transition allows the estimator to shift smoothly between explicit grounding and LLM-based judgment. The parameters $\gamma$ and $\lambda$ are tuned on a held-out validation set to minimize the discrepancy between predicted hallucination scores and reference annotations. The combined estimator satisfies:
\begin{itemize}
    \item \textbf{Boundedness:} $\mathcal{H}(t,k) \in [0,1]$.
    \item \textbf{Adaptivity:} the score changes according to grounding reliability.
    \item \textbf{Robustness:} the estimator reduces dependence on either component alone.
\end{itemize}
By integrating reference grounding with LLM-based semantic judgment, the estimator remains applicable across domains with different evidence availability. This is essential in multi-agent cascades, where intermediate outputs vary in factual support, semantic coherence, and susceptibility to propagated hallucination. Furthermore, let $\mathcal{C}(t)=\{c_1,\ldots,c_n\}$ denote the claims extracted from text $t$. Rule-based verification estimates each claim's factual support using the reference knowledge base $\mathcal{G}_k$ and penalizes recurring hallucination patterns that semantic similarity alone does not capture. For each claim $c_j$, the grounding score is:
\begin{equation}
s(c_j)=
\max_{g \in \mathcal{G}_k}
\frac{\mathbf{e}_{c_j}\cdot \mathbf{e}_g}
{\|\mathbf{e}_{c_j}\|\|\mathbf{e}_g\|}
\end{equation}
where $\mathbf{e}_{c_j}$ and $\mathbf{e}_g$ denote the embeddings of claim $c_j$ and reference fact $g$. Higher values indicate alignment with verified knowledge, while lower values indicate weak and missing factual support.
To capture recurrent hallucination signatures, we define:
\begin{equation}
\psi(c_j)=
\max_{h \in \mathcal{P}_k}
\sigma\left(\mathbf{e}_{c_j}^{\top}\mathbf{e}_h\right)
\end{equation}
where $\mathcal{P}_k$ is a domain-specific set of hallucination patterns and $\sigma(x)=(1+e^{-x})^{-1}$. This term captures fabricated entities, unsupported relations, temporal inconsistencies, and logical contradictions.
The claim-level rule-based hallucination score is:
\begin{equation}
r(c_j,k)=
\frac{(1-s(c_j))+\lambda \cdot \psi(c_j)}
{1+\lambda}
\end{equation}
where $\lambda \geq 0$ controls the contribution of pattern-based penalties. The response-level score is:
\begin{equation}
\mathcal{H}_{\text{rule}}(t,k)=
\frac{1}{|\mathcal{C}(t)|}
\sum_{c_j \in \mathcal{C}(t)} r(c_j,k).
\end{equation}
Since $s(c_j)$ and $\psi(c_j)$ are bounded in $[0,1]$, the normalized score satisfies:
\begin{equation}
\mathcal{H}_{\text{rule}}(t,k) \in [0,1].
\end{equation}
Thus, $\mathcal{H}_{\text{rule}}(t,k)$ provides an interpretable measure of factual inconsistency by combining semantic grounding with hallucination-pattern activation. Additionally, rule-based verification relies on explicit grounding and may miss plausible but unsupported claims. To address this limitation, we use an LLM-based judge to estimate claim-level hallucination from semantic and contextual evidence.
For each claim $c_j \in \mathcal{C}(t)$ and surrounding context $c$, the contextual representation is:
\begin{equation}
\phi(c_j,c) \in \mathbb{R}^d
\end{equation}
computed by encoding the claim-context pair with an instruction-tuned LLM:
\begin{equation}
\phi(c_j,c)=\mathrm{Enc}_{\text{LLM}}([c_j \oplus c]).
\end{equation}
This representation captures the claim's semantics and its consistency with the surrounding response.
The judge classifies each claim into $K$ factuality categories, including factual correctness, unsupported inference, factual fabrication, logical contradiction, and contextual inconsistency. Given $\phi(c_j,c)$, it produces logits:
\begin{equation}
\mathbf{z}_j=\mathbf{W}\phi(c_j,c)+\mathbf{b}, 
\quad 
\mathbf{z}_j \in \mathbb{R}^K
\end{equation}
which are converted into categorical probabilities:
\begin{equation}
\boldsymbol{\pi}_j=\mathrm{softmax}(\mathbf{z}_j).
\end{equation}
The judge follows a structured prompt:
\begin{quote}
\textit{Given the following claim and context, determine whether the claim is factually supported. Classify the claim into one of the following categories: 1) factually correct, 2) unsupported inference, 3) fabricated fact, 4) logical contradiction, 5) contextual inconsistency. Provide a probability distribution over the categories.}
\end{quote}
The categorical probabilities are mapped to a scalar hallucination probability:
\begin{equation}
q_j=\sigma\left(\mathbf{w}^{\top}\boldsymbol{\pi}_j\right)
\end{equation}
where $\mathbf{w}$ assigns lower weight to factual correctness and higher weights to hallucination-related categories. The response-level model-based hallucination score is:
\begin{equation}
\mathcal{H}_{\text{LLM}}(t,k)=
\frac{1}{|\mathcal{C}(t)|}
\sum_{c_j \in \mathcal{C}(t)} q_j.
\end{equation}
To improve probabilistic interpretability, the estimator is calibrated so that predicted hallucination probabilities align with empirical annotations:
\begin{equation}
\mathbb{E}[q_j \mid y_j=1] \approx 1,
\quad
\mathbb{E}[q_j \mid y_j=0] \approx 0
\end{equation}
where $y_j=1$ denotes a hallucinated claim and $y_j=0$ denotes a supported claim. Calibration uses temperature scaling:
\begin{equation}
\tilde{\boldsymbol{\pi}}_j=
\mathrm{softmax}\left(\frac{\mathbf{z}_j}{\tau}\right)
\end{equation}
where $\tau>0$ controls prediction sharpness and is optimized on a held-out validation set by minimizing expected calibration error. This calibration keeps hallucination scores comparable across agents, cascade depths, cross-model configurations, injected hallucination experiments, and long-chain evaluations.
In the sequential setting, each agent conditions on the original prompt and the accumulated outputs of preceding agents:
\begin{equation}
o_i \sim P_{M_i}(\cdot \mid p,o_1,\ldots,o_{i-1}), 
\quad i=1,\ldots,L
\end{equation}
or equivalently:
\begin{equation}
o_i =
M_i\left(p \oplus \left(\bigoplus_{j=1}^{i-1} o_j\right)\right)
\end{equation}
where $\oplus$ denotes concatenation. This accumulated context enables iterative refinement but also creates a pathway for earlier errors to impact later agents. Thus, each agent may attenuate, preserve, transform, amplify, and recover hallucinated content from previous outputs.
Let $h_i=\mathcal{H}(o_i)$ denote the hallucination score at step $i$. The sequence $\{h_i\}_{i=1}^{L}$ defines the hallucination trajectory, modeled as:
\begin{equation}
h_i = \alpha_i h_{i-1} + \beta_i \mathcal{H}(p) + \epsilon_i
\end{equation}
where $\alpha_i$ captures inter-agent propagation, $\beta_i$ models prompt-level uncertainty, and $\epsilon_i \sim \mathcal{N}(0,\sigma_i^2)$ represents stochastic generation variability. This induces the first-order dependency:
\begin{equation}
P(h_i \mid h_1,\ldots,h_{i-1}) = P(h_i \mid h_{i-1}).
\end{equation}
The cascade is stable when:
\begin{equation}
|\alpha_i| < 1 \quad \forall i
\end{equation}
ensuring bounded error propagation. Under constant coefficients $(\alpha_i=\alpha,\beta_i=\beta)$, the trajectory admits:
\begin{equation}
h_i =
\alpha^{i-1}h_1
+
\beta \mathcal{H}(p)
\sum_{j=0}^{i-2}\alpha^j
+
\sum_{j=0}^{i-2}\alpha^j\epsilon_{i-j}
\end{equation}
and, if $|\alpha|<1$, converges to:
\begin{equation}
\lim_{i \to \infty} h_i =
\frac{\beta \mathcal{H}(p)}{1-\alpha}.
\end{equation}
This supports long-chain analysis by identifying whether additional agents reduce hallucination, reach saturation, introduce drift, and trigger failure cases.
Propagation is quantified using:
\begin{equation}
\rho_i =
\frac{h_i-h_{i-1}}{h_{i-1}+\epsilon},
\quad
\Gamma(L)=h_L-h_1
\end{equation}
and:
\begin{equation}
\mathcal{A}_L =
\frac{h_L}{h_1+\epsilon},
\quad
\epsilon=10^{-8}.
\end{equation}
Here, $\rho_i$ captures step-wise propagation, $\Gamma(L)$ measures total hallucination change, and $\mathcal{A}_L$ summarizes global attenuation and amplification. Values of $\mathcal{A}_L<1$ indicate net attenuation, values near one indicate preservation, and values above one indicate amplification.
In the parallel setting, $N$ agents independently generate outputs from the same prompt:
\begin{equation}
o_j \sim P_{M_j}(\cdot \mid p), \quad j=1,\ldots,N .
\end{equation}
In contrast to sequential cascades, this architecture removes dependency on previous outputs and uses inter-agent disagreement to identify unreliable claims. Let $h_j=\mathcal{H}(o_j)$ denote the hallucination score of agent $j$. The final output is selected using a similarity-weighted reliability objective:
\begin{equation}
o^* =
\arg\max_{o_j}
\sum_{m=1}^{N}
w_m \cdot \mathrm{sim}(o_j,o_m)\cdot (1-h_m)
\end{equation}
where $\mathrm{sim}(o_j,o_m)$ measures semantic agreement and $(1-h_m)$ weights each output by factual reliability.
The model reliability weights are:
\begin{equation}
w_m =
\frac{\exp(-\gamma \cdot \mathcal{H}_{\text{base}}(M_m))}
{\sum_{n=1}^{N}\exp(-\gamma \cdot \mathcal{H}_{\text{base}}(M_n))}
\end{equation}
where $\gamma>0$ controls sensitivity to model reliability, and $\mathcal{H}_{\text{base}}(M_m)$ is the baseline hallucination level of model $M_m$ estimated from validation data. This weighting favors historically reliable models while preserving heterogeneous generation.
The hallucination distribution across agents is summarized by:
\begin{equation}
\mathcal{H}_{\min}=\min_j h_j,
\quad
\mathcal{H}_{\text{mean}}=
\frac{1}{N}\sum_{j=1}^{N}h_j
\end{equation}
and:
\begin{equation}
\mathcal{H}_{\text{var}}=
\frac{1}{N}
\sum_{j=1}^{N}
(h_j-\mathcal{H}_{\text{mean}})^2 .
\end{equation}
Here, $\mathcal{H}_{\min}$ captures the best-case response, $\mathcal{H}_{\text{mean}}$ reflects average hallucination, and $\mathcal{H}_{\text{var}}$ measures inter-agent diversity. High variance indicates useful disagreement for consensus selection, whereas low variance suggests correlated failures.
The gain from parallel aggregation is:
\begin{equation}
\Delta =
\mathcal{H}_{\min}^{\text{single}}
-
\mathcal{H}_{\min}^{\text{ensemble}} .
\end{equation}
A positive $\Delta$ indicates that model diversity reduces hallucination beyond individual model performance. Thus, parallel consensus complements sequential propagation by leveraging heterogeneous outputs to reduce correlated factual errors before downstream refinement.
Let $\{S_1,\ldots,S_K\}$ denote a set of specialist agents, where each specialist targets a specific task dimension, such as factual retrieval, analytical reasoning, and domain-specific knowledge. Each specialist generates:
\begin{equation}
s_i \sim P_{S_i}(\cdot \mid p), \quad i=1,\ldots,K .
\end{equation}
Specialists are instantiated through heterogeneous prompting strategies applied to the same underlying LLM, increasing output diversity and reducing correlated hallucinations across reasoning perspectives.
Let $h_i=\mathcal{H}(s_i)$ denote the hallucination score of specialist $i$. A coordinator aggregates specialist outputs as:
\begin{equation}
o \sim P_C(\cdot \mid p,s_1,\ldots,s_K).
\end{equation}
The coordinator is conditioned on the concatenated specialist outputs and guided to prioritize factual consistency, semantic coherence, and preservation of verified information.
Let $\mathbf{e}_{s_i}$ and $\mathbf{e}_p$ denote embeddings of specialist output $s_i$ and prompt $p$. The attention weights are:
\begin{equation}
a_i =
\frac{\exp(u_i)}
{\sum_{j=1}^{K}\exp(u_j)},
\quad
u_i =
\mathbf{v}^{\top}
\tanh(\mathbf{W}_s \mathbf{e}_{s_i}+\mathbf{W}_p \mathbf{e}_p)
\end{equation}
and the weighted representation is:
\begin{equation}
\tilde{s} =
\sum_{i=1}^{K} a_i \mathbf{e}_{s_i}.
\end{equation}
To incorporate factual reliability, attention is modulated by hallucination scores:
\begin{equation}
a_i \propto \exp(-\lambda h_i)
\end{equation}
where $\lambda>0$ controls sensitivity to hallucination. This favors low-hallucination specialists and down-weights unreliable outputs.
The overall hierarchical hallucination score is:
\begin{equation}
\mathcal{H}_{\text{hier}} =
\beta \cdot \frac{1}{K}\sum_{i=1}^{K}h_i
+
(1-\beta)\cdot \mathcal{H}(o)
\end{equation}
where $\beta \in [0,1]$ balances specialist-level reliability and final coordinator quality. The optimal value is:
\begin{equation}
\beta^* =
\arg\min_{\beta \in [0,1]}
\mathbb{E}_{p \sim \mathcal{P}}
\left[
\mathcal{H}_{\text{hier}}(p,\beta)
\right].
\end{equation}
In practice, $\beta$ is tuned on a held-out validation set. This architecture supports reliability-aware information flow by amplifying low-hallucination specialist outputs and suppressing high-risk outputs. It also supports failure analysis by identifying whether hallucinations originate from individual specialists, coordinator synthesis, cross-specialist inconsistencies, and downstream propagation.

\subsection{Hallucination Cascade Dynamics}
\label{sec:cascade_dynamics}
Let $h_i=\mathcal{H}(o_i)$ denote the hallucination score at cascade step $i$. The sequence $\{h_i\}$ captures the temporal evolution of hallucination as information passes across agents. We model this evolution as a linear stochastic process with prompt-level input:
\begin{equation}
h_i =
\sum_{r=1}^{R}\lambda_r h_{i-r}
+
\mu \mathcal{H}(p)
+
\eta_i
\end{equation}
where $\lambda_r$ captures the impact of previous hallucination states, $\mu$ measures the contribution of the original prompt, and $\eta_i \sim \mathcal{N}(0,\sigma^2)$ represents stochastic generation variability. Larger $\lambda_r$ values indicate persistence of prior factual errors, while smaller values indicate attenuation across later agents.
Defining the state vector as:
\begin{equation}
\mathbf{x}_i =
[h_{i-1},\ldots,h_{i-R},\mathcal{H}(p)]^{T}
\end{equation}
The process can be written compactly as:
\begin{equation}
h_i =
\mathbf{\Phi}^{T}\mathbf{x}_i+\eta_i
\end{equation}
with:
\begin{equation}
\mathbf{\Phi} =
[\lambda_1,\ldots,\lambda_R,\mu]^{T}.
\end{equation}
This representation supports time-series analysis, parameter estimation, stability assessment, and trajectory prediction. It also enables comparison across cross-model and long-chain cascades through their estimated propagation coefficients. Thus, hallucination is modeled as a structured temporal process in which factual errors may persist, attenuate, recover, amplify, and trigger failure cases across agent interactions. Hallucination dynamics are modeled as an autoregressive process:
\begin{equation}
h_i =
\sum_{r=1}^{R}\lambda_r h_{i-r}
+
\mu \mathcal{H}(p)
+
\eta_i
\end{equation}
where previous hallucination states, prompt-level uncertainty, and stochastic generation noise jointly determine the hallucination score at step $i$. The process is stable if the characteristic polynomial
\begin{equation}
z^R-\lambda_1 z^{R-1}-\cdots-\lambda_R=0
\end{equation}
has all roots inside the unit circle, ensuring that past hallucination states decay rather than accumulate.
For the first-order case:
\begin{equation}
h_i =
\alpha h_{i-1}
+
\mu \mathcal{H}(p)
+
\eta_i
\end{equation}
stability reduces to:
\begin{equation}
|\alpha|<1
\end{equation}
where smaller $\alpha$ values indicate attenuation and values near one indicate persistence. Ignoring stochastic noise, the solution is:
\begin{equation}
h_i =
\alpha^{i-1}h_1
+
\mu \mathcal{H}(p)
\cdot
\frac{1-\alpha^{i-1}}{1-\alpha}.
\end{equation}
If $|\alpha|<1$, the process converges to:
\begin{equation}
\lim_{i \to \infty} h_i =
\frac{\mu \mathcal{H}(p)}{1-\alpha}.
\end{equation}
Thus, hallucination may persist even under stable dynamics when the original prompt is ambiguous and lacks factual context.
Local and global propagation are measured as:
\begin{equation}
\gamma_i=h_i-h_{i-1},
\quad
\Gamma(L)=h_L-h_1
\end{equation}
and:
\begin{equation}
\rho_i=
\frac{h_i-h_{i-1}}{h_{i-1}+\epsilon}.
\end{equation}
Here, $\gamma_i$ captures absolute step-wise change, $\Gamma(L)$ measures total cascade-level deviation, and $\rho_i$ measures normalized propagation. Negative values indicate attenuation, positive values indicate amplification, and values near zero indicate preservation.
Under the first-order form with $\delta=\mu\mathcal{H}(p)$:
\begin{equation}
h_i=
\alpha h_{i-1}+\delta
\end{equation}
The dominant behavior can be approximated as:
\begin{equation}
h_i \approx h_1 e^{\kappa(i-1)},
\quad
\kappa=\log \alpha .
\end{equation}
Values of $\kappa<0$ indicate attenuation, while $\kappa>0$ indicates amplification, supporting long-chain analysis of saturation, suppression, and renewed error growth.
Let $\mathcal{I}(t)$ denote semantic information preserved in text $t$. Its evolution across the cascade is:
\begin{equation}
\mathcal{I}(o_i)=
\mathcal{I}(p)
\prod_{j=1}^{i}(1-h_j).
\end{equation}
For small $h_j$:
\begin{equation}
\mathcal{I}(o_i)
\approx
\mathcal{I}(p)
\exp\left(-\sum_{j=1}^{i}h_j\right).
\end{equation}
Defining $\lambda=\frac{1}{i}\sum_{j=1}^{i}h_j$, we obtain:
\begin{equation}
\mathcal{I}(o_i)
\approx
\mathcal{I}(p)e^{-\lambda i}.
\end{equation}
This formulation shows that hallucination can act as an information-dissipation factor. Even when hallucination decreases, repeated refinement may compress, weaken, and transform factual content. This supports the analysis of semantic drift, recovery patterns, and long-chain failure cases.
The cascade is implemented as a sequential multi-agent generation process. Each agent $A_i$ produces an output $o_i$ conditioned on the original prompt and the accumulated outputs from preceding agents. At each step, the hallucination estimator computes $h_i=\mathcal{H}(o_i)$, forming the trajectory $\mathbf{h}=(h_1,\ldots,h_L)$. The relative propagation factor $\rho_i$ measures step-wise change between consecutive agents, where $\rho_i<0$ indicates attenuation, $\rho_i\approx0$ indicates preservation, and $\rho_i>0$ indicates amplification. The cumulative deviation $\Gamma(L)$ measures the total change from the initial response, while the amplification factor $\mathcal{A}_L$ summarizes global cascade behavior. Additionally, these quantities support the analysis of propagation, long-chain stability, recovery of injected hallucinations, and failure dynamics.
\begin{algorithm}[t]
\footnotesize
\setlength{\abovedisplayskip}{3pt}
\setlength{\belowdisplayskip}{3pt}
\caption{Hallucination Propagation in a Multi-Agent Cascade}
\label{alg:cascade_ieee}
\begin{algorithmic}[1]
\Require Prompt $p$, agents $\mathcal{C}=(A_1,\dots,A_L)$, estimator $\mathcal{H}(\cdot)$
\Ensure Trajectory $\mathbf{h}=(h_1,\dots,h_L)$, propagation $\{\rho_i\}_{i=2}^{L}$, $\Gamma(L)$, $\mathcal{A}_L$

\State $o_1 \gets A_1(p)$ \Comment{Initial generation}
\State $h_1 \gets \mathcal{H}(o_1)$ \Comment{Initial hallucination score}

\For{$i=2$ to $L$}
    \State $x_i \gets p \oplus \left(\bigoplus_{j=1}^{i-1} o_j\right)$ 
    \Comment{Aggregate prior context}
    
    \State $o_i \gets A_i(x_i)$ 
    \Comment{Generate refined output}
    
    \State $h_i \gets \mathcal{H}(o_i)$ 
    \Comment{Estimate hallucination}
    
    \State $\rho_i \gets \dfrac{h_i-h_{i-1}}{\max(h_{i-1},\varepsilon)}$ 
    \Comment{Compute relative propagation}
\EndFor

\State $\Gamma(L) \gets h_L-h_1$ 
\Comment{Compute cumulative deviation}

\If{$h_1>0$}
    \State $\mathcal{A}_L \gets \dfrac{h_L}{h_1}$ 
    \Comment{Compute amplification factor}
\Else
    \State $\mathcal{A}_L \gets h_L$
    \Comment{Handle zero initial hallucination}
\EndIf

\State \Return $\mathbf{h},\{\rho_i\}_{i=2}^{L},\Gamma(L),\mathcal{A}_L$
\end{algorithmic}
\end{algorithm}

\section{Experimental Setup}
\label{sec:experimental_setup}
This section describes the experimental design used to evaluate hallucination as a dynamic process in multi-agent LLM systems. The setup links the proposed formulation to measurable cascade behavior, including propagation, attenuation, recovery, semantic drift, long-chain stability, and failure dynamics.

\subsection{Multi-Agent Experimental Design for Hallucination Analysis}
\label{sec:multi_agent_experimental_design}
The experimental design evaluates how hallucination evolves across sequential agents, how cascade depth impacts factual reliability, and how model heterogeneity impacts correction behavior. Three LLMs are used: GPT-5.3, DeepSeek-V3, and LLaMA-3-70B-Instruct. These models are selected to capture different reliability, generation, and efficiency profiles. Table~\ref{tab:exp_setup} summarizes the experimental configuration. The core benchmark uses sequential cascades in which each downstream agent receives the original prompt together with the accumulated outputs from all preceding agents. This design makes hallucination observable as a trajectory across agents rather than as a single response-level score. The baseline benchmark consists of 500 cascade runs, equally divided between two cascade depths: 250 runs with 2-agent chains and 250 runs with 3-agent chains. Because each agent output is evaluated independently, the 2-agent cascades produce 500 evaluated agent-level responses, while the 3-agent cascades produce 750 evaluated agent-level responses. Therefore, the baseline benchmark contains 1,250 total evaluated responses. The analysis also includes cross-model cascades, long-chain cascades, injected-hallucination cascades, error-recovery analysis, and failure-case analysis. These settings are designed to evaluate correlated hallucination reduction, convergence, saturation, rebound impacts, information loss, controlled error propagation, correction behavior, overcorrection, semantic drift, and unresolved hallucination. The benchmark spans 10 knowledge domains: Roman Empire, Photosynthesis, DNA, Quantum Computing, Vaccines, Climate Change, Black Holes, CRISPR, Blockchain, and Machine Learning. Each domain is evaluated across 50 baseline runs. All experiments use consistent prompting and inference settings. Responses are decomposed into atomic claims, evaluated through rule-based grounding and LLM-based semantic judgment, and converted into normalized hallucination scores in $[0,1]$. In addition to hallucination, the analysis measures factual accuracy, response quality, semantic drift, cascade risk, response time, token usage, inference cost, and propagation metrics.
\begin{table*}[t]
\centering
\footnotesize
\setlength{\tabcolsep}{4pt}
\renewcommand{\arraystretch}{1.15}
\caption{Summary of the experimental configuration for hallucination propagation analysis in multi-agent LLM cascades.}
\label{tab:exp_setup}
\resizebox{0.92\textwidth}{!}{
\begin{tabular}{p{0.24\textwidth} p{0.64\textwidth}}
\toprule
\textbf{Component} & \textbf{Configuration} \\
\midrule
Models & GPT-5.3, DeepSeek-V3, LLaMA-3-70B-Instruct \\
Cascade Structures & Sequential cascades, cross-model cascades, long-chain cascades, injected hallucination cascades \\
Baseline Chain Lengths & 2-agent and 3-agent chains \\
Extended Chain Lengths & 5-agent and 7-agent chains for convergence, saturation, rebound, and failure analysis \\
Baseline Cascade Runs & 500 total cascade runs \\
Runs per Baseline Chain Length & 250 runs with 2-agent cascades and 250 runs with 3-agent cascades \\
Evaluated Outputs from 2-Agent Cascades & 500 agent-level responses \\
Evaluated Outputs from 3-Agent Cascades & 750 agent-level responses \\
Total Baseline Evaluated Outputs & 1,250 agent-level responses \\
Knowledge Domains & Roman Empire, Photosynthesis, DNA, Quantum Computing, Vaccines, Climate Change, Black Holes, CRISPR, Blockchain, Machine Learning \\
Runs per Domain & 50 baseline runs per domain \\
Core Metrics & Hallucination, factual accuracy, response quality, semantic drift, cascade risk, response time, token usage, inference cost \\
Propagation Metrics & Hallucination trajectory, relative propagation, cumulative deviation, amplification factor, recovery rate, failure indicator \\
Logged Artifacts & Prompts, model outputs, extracted claims, grounding scores, judge scores, response-level metrics, transition-level metrics \\
\bottomrule
\end{tabular}
}
\end{table*}

\subsection{Cascade Configurations}
\label{sec:cascade_configurations}
The cascade configurations isolate the impact of interaction structure on hallucination behavior. As shown in Table~\ref{tab:cascade_configurations}, short sequential cascades provide the baseline for attenuation and amplification analysis. Cross-model cascades evaluate heterogeneous model sequences; long-chain cascades assess convergence and degradation with added depth; and injected-hallucination cascades trace controlled factual errors across agents. Error recovery and failure-case analyses connect aggregate metrics with claim-level outcomes.
\begin{table*}[t]
\centering
\footnotesize
\setlength{\tabcolsep}{4pt}
\renewcommand{\arraystretch}{1.15}
\caption{Cascade configurations used to evaluate hallucination propagation, recovery, and failure dynamics.}
\label{tab:cascade_configurations}
\resizebox{0.95\textwidth}{!}{
\begin{tabular}{p{0.20\textwidth} p{0.30\textwidth} p{0.38\textwidth}}
\toprule
\textbf{Configuration} & \textbf{Purpose} & \textbf{Measured Behavior} \\
\midrule
Short Sequential Cascades & Evaluate baseline propagation across 2-agent and 3-agent chains & Attenuation, amplification, factual decay, semantic drift \\
Cross-Model Cascades & Evaluate heterogeneous sequences across GPT-5.3, DeepSeek-V3, and LLaMA-3-70B-Instruct & Model-dependent correction, correlated hallucination reduction, sequence impacts \\
Long-Chain Cascades & Extend cascade depth beyond the baseline setting & Convergence, saturation, rebound, long-chain stability, information loss \\
Injected Hallucination Cascades & Insert controlled factual errors into early-stage outputs & Error persistence, correction probability, transformation, amplification \\
Error Recovery Analysis & Track hallucinated claims across consecutive agents & Corrected claims, preserved claims, weakened claims, amplified claims \\
Failure Case Analysis & Examine representative unsuccessful trajectories & Catastrophic propagation, overcorrection, semantic collapse, unresolved hallucination \\
\bottomrule
\end{tabular}
}
\end{table*}

\subsection{Evaluation Metrics}
\label{sec:evaluation_metrics}
The evaluation uses three metric groups: response-level reliability, efficiency, and propagation behavior. Hallucination is measured using the proposed claim-level estimator, while factual accuracy, response quality, semantic drift, and cascade risk assess factual correctness, utility, meaning shift, and reliability degradation, respectively. Efficiency is measured using response time, token usage, and inference cost.
For propagation analysis, the hallucination trajectory $\mathbf{h}$ tracks factual inconsistency across agents. Table~\ref{tab:evaluation_metrics} summarizes the evaluation metrics used to quantify hallucination propagation, correction, and failure. The relative propagation factor $\rho_i$, cumulative deviation $\Gamma(L)$, and amplification factor $\mathcal{A}_L$ quantify local, cumulative, and global hallucination change, respectively. The recovery rate captures corrected hallucinated claims, whereas the failure indicator identifies unresolved and amplified hallucination trajectories.
\begin{table*}[t]
\centering
\footnotesize
\setlength{\tabcolsep}{4pt}
\renewcommand{\arraystretch}{1.15}
\caption{Evaluation metrics used for response-level assessment and hallucination propagation analysis.}
\label{tab:evaluation_metrics}
\resizebox{0.95\textwidth}{!}{
\begin{tabular}{p{0.18\textwidth} p{0.40\textwidth} p{0.32\textwidth}}
\toprule
\textbf{Metric} & \textbf{Definition} & \textbf{Interpretation} \\
\midrule
Hallucination Score & Normalized claim-level factual inconsistency estimated through hybrid grounding and LLM-based judgment & Lower values indicate factual reliability \\
Factual Accuracy & Agreement between generated claims and verified reference facts & Higher values indicate factual correctness \\
Response Quality & Coherence, completeness, and usefulness of the generated response & Higher values indicate better response utility \\
Semantic Drift & Deviation from the original prompt and preserved factual content & Higher values indicate greater meaning shift \\
Cascade Risk & Combined reliability risk based on hallucination, drift, and factual degradation & Higher values indicate less stable cascade behavior \\
Response Time & Wall-clock generation time per agent output & Lower values indicate higher efficiency \\
Token Usage & Number of input and output tokens consumed during generation & Lower values indicate lower computational burden \\
Inference Cost & Estimated generation cost based on token usage & Lower values indicate lower operational cost \\
Trajectory $\mathbf{h}$ & Sequence of hallucination scores across agents & Shows temporal evolution of hallucination \\
Relative Propagation $\rho_i$ & Normalized hallucination change between consecutive agents & Negative values indicate attenuation; positive values indicate amplification \\
Cumulative Deviation $\Gamma(L)$ & Difference between final and initial hallucination scores & Measures total cascade-level change \\
Amplification Factor $\mathcal{A}_L$ & Ratio between final and initial hallucination scores & Values below one indicate net attenuation \\
Recovery Rate & Fraction of hallucinated claims corrected in later stages & Higher values indicate correction ability \\
Failure Indicator & Marker for unresolved, amplified, and semantically damaging hallucination trajectories & Identifies representative failure cases \\
\bottomrule
\end{tabular}
}
\end{table*}

\subsection{Baselines}
\label{sec:baselines}
The proposed cascade is compared with four baselines, summarized in Table~\ref{tab:baselines}. These baselines represent direct generation, reasoning-based prompting, retrieval-grounded generation, and same-model iterative revision. All baselines use the same prompts, knowledge domains, and evaluation metrics as the proposed method, ensuring that performance differences reflect the impact of multi-agent interaction rather than variations in task design and evaluation procedure.
\begin{table*}[t]
\centering
\footnotesize
\setlength{\tabcolsep}{3pt}
\renewcommand{\arraystretch}{1.05}
\caption{Baseline methods used for comparison in the experimental setup.}
\label{tab:baselines}
\resizebox{0.85\textwidth}{!}{
\begin{tabular}{p{0.22\textwidth} p{0.58\textwidth}}
\toprule
\textbf{Baseline} & \textbf{Description} \\
\midrule
Single-Agent & Direct response generation from the prompt without refinement \\
CoT & Step-by-step reasoning before final answer generation \\
RAG & Response generation conditioned on retrieved external documents \\
Self-Refinement & Same-model iterative revision using the original prompt and prior output \\
\bottomrule
\end{tabular}
}
\end{table*}

\subsection{Research Questions}
\label{sec:research_questions}
This study is guided by three research questions that examine hallucination as a dynamic process in multi-agent LLM cascades:

\begin{itemize}
    \item \textbf{RQ1: How does hallucination evolve across sequential multi-agent LLM cascades?} \\
    This question analyzes whether hallucination accumulates, attenuates, recovers, and fluctuates as responses pass through multiple agents. It is examined using hallucination trajectories, propagation rates, cumulative deviation, amplification factors, recovery patterns, injected hallucination behavior, and failure cases.

    \item \textbf{RQ2: How does cascade depth impact hallucination behavior and semantic quality?} \\
    This question evaluates whether deeper cascades improve hallucination mitigation and introduce saturation, semantic drift, factual decay, information loss, and degradation of response quality. It is addressed by comparing short and long-chain cascades.

    \item \textbf{RQ3: How do different LLMs and cross-model cascade structures compare in hallucination propagation, output quality, latency, and cost?} \\
    This question compares GPT-5.3, DeepSeek-V3, and LLaMA-3-70B-Instruct across homogeneous and heterogeneous cascades to assess factual reliability, hallucination attenuation, recovery, semantic drift, response quality, latency, token usage, and inference cost.
\end{itemize}

\section{Experimental Results and Analysis}
\label{Experimental Results and Analysis}
This section summarizes the empirical findings obtained from the evaluation.

\subsection{Overall System and Model Performance}
\label{sec:statistical_analysis}
\begin{figure*}[t]
\centering
\includegraphics[width=0.80\textwidth]{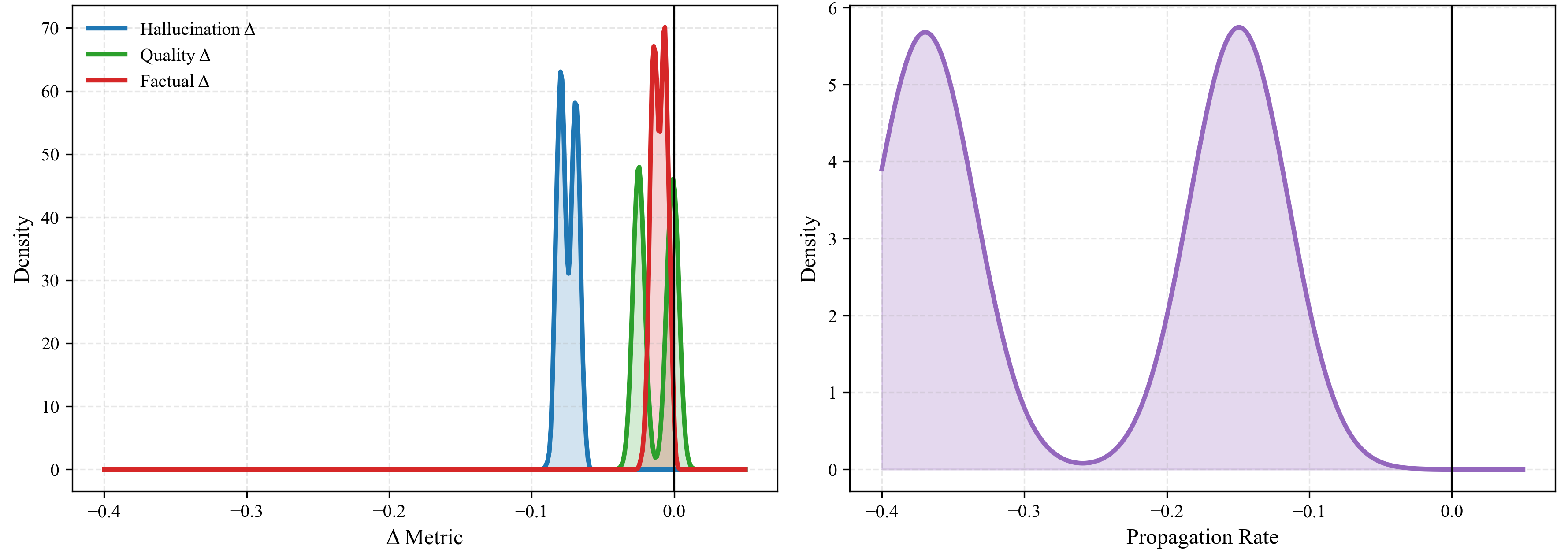}
\caption{Cost-performance trade-off landscape across evaluated LLMs, illustrating relationships between hallucination, accuracy, response quality, latency, token usage, and computational cost.}
\label{fig:cost}
\end{figure*}

\begin{figure*}[t]
\centering
\includegraphics[width=\textwidth]{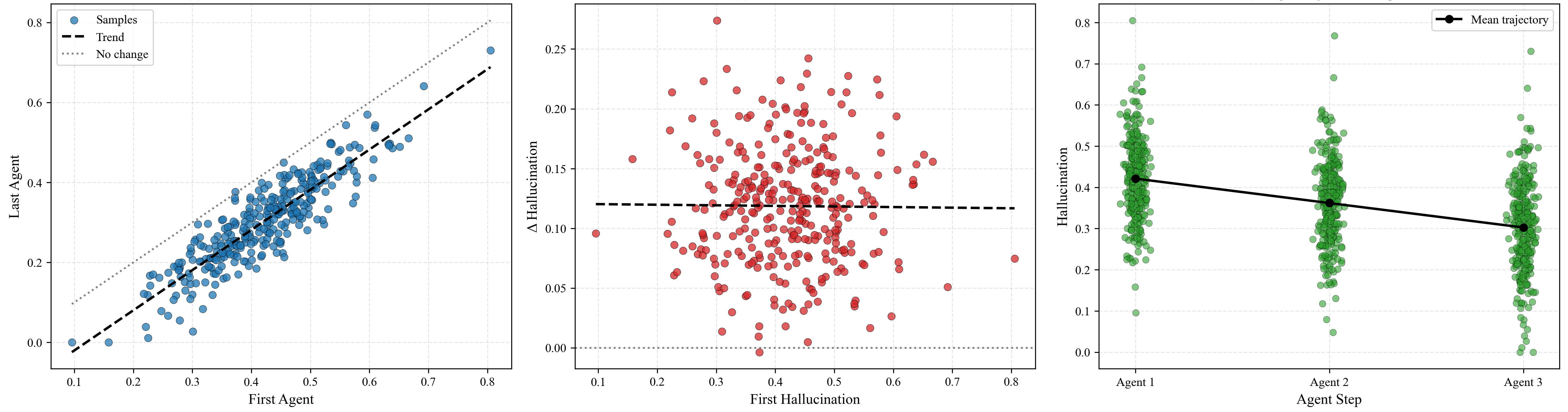}
\caption{Density profiles of LLM performance across models, illustrating the distribution of hallucination, accuracy, semantic drift, response quality, latency, token usage, and cost.}
\label{fig:density}
\end{figure*}

To address RQ1 and RQ3, we analyze aggregate system behavior and model-level reliability--efficiency trade-offs. Table~\ref{tab:overall_summary} summarizes results across models, cascade configurations, and domains. The system produced 1,250 successfully evaluated outputs across 500 experiments with no failed rows, indicating stable execution. The average hallucination score was 0.360914 with a standard deviation of 0.118079, and the 95\% confidence interval was:
\begin{equation}
\text{CI}_{95\%} \approx 0.361 \pm 1.96 \cdot \frac{0.118}{\sqrt{1250}} \approx [0.354, 0.368].
\end{equation}
This narrow interval indicates that the average hallucination level is stable. The system also maintains relatively high factual accuracy (0.785320) and response quality (0.748253), while showing measurable semantic drift (0.262361) and cascade risk (0.314116). These results suggest that hallucination interacts with factuality, quality, drift, and risk in a structured way rather than behaving as random noise.
\begin{table}[t]
\centering
\footnotesize
\setlength{\tabcolsep}{3pt}
\renewcommand{\arraystretch}{0.95}
\caption{Overall experimental summary, aggregating results across models, cascade configurations, and knowledge domains.}
\label{tab:overall_summary}
\resizebox{0.3\textwidth}{!}{
\begin{tabular}{l r}
\toprule
Metric & Value \\
\midrule
Total rows & 1250 \\
Total experiments & 500 \\
Successful rows & 1250 \\
Failed rows & 0 \\
Average hallucination & 0.360914 \\
Hallucination std. dev. & 0.118079 \\
Average factual accuracy & 0.785320 \\
Average quality & 0.748253 \\
Average cascade risk & 0.314116 \\
Average semantic drift & 0.262361 \\
Average response time (s) & 10.470044 \\
Total tokens & 869801 \\
Total cost & 0.869801 \\
\bottomrule
\end{tabular}
}
\end{table}
Table~\ref{tab:model_performance} shows a clear reliability--efficiency trade-off. LLaMA-3-70B-Instruct achieves the lowest hallucination score (0.272413), but also has the highest response time (15.082857 s). GPT-5.3 is the fastest and least costly model, with a response time of 5.248978 s and a cost of 0.000491, but it has the highest hallucination score (0.417564). DeepSeek-V3 falls between the two models in terms of reliability, latency, and cost.
\begin{table*}[t]
\centering
\footnotesize
\setlength{\tabcolsep}{3pt}
\renewcommand{\arraystretch}{0.95}
\caption{Model-level performance comparison across metrics, including hallucination, accuracy, response quality, semantic drift, latency, token usage, and cost.}
\label{tab:model_performance}
\resizebox{0.6\textwidth}{!}{
\begin{tabular}{l r r r r r r r}
\toprule
Model & Rows & Halluc. & Factual & Quality & Time (s) & Tokens & Cost \\
\midrule
GPT-5.3 & 500 & 0.417564 & 0.792063 & 0.763400 & 5.248978 & 491.496 & 0.000491 \\
DeepSeek-V3 & 500 & 0.348514 & 0.786664 & 0.738763 & 13.384704 & 840.748 & 0.000841 \\
LLaMA-3-70B-Instruct & 250 & 0.272413 & 0.769147 & 0.736940 & 15.082857 & 814.716 & 0.000815 \\
\bottomrule
\end{tabular}
}
\end{table*}
The ranking in Table~\ref{tab:model_ranking} confirms that reliability and efficiency do not align. LLaMA-3-70B-Instruct ranks first in hallucination reduction, while GPT-5.3 is the most efficient in terms of time and cost. Thus, model selection should depend on the deployment objective: high-reliability settings may favor lower hallucination, whereas latency-sensitive settings may require faster models with additional verification.
\begin{table}[t]
\centering
\footnotesize
\setlength{\tabcolsep}{3pt}
\renewcommand{\arraystretch}{0.95}
\caption{Model ranking by reliability and efficiency, comparing LLMs based on hallucination, accuracy, response quality, latency, token usage, and computational cost.}
\label{tab:model_ranking}
\resizebox{0.5\textwidth}{!}{
\begin{tabular}{r l r r r}
\toprule
Rank & Model & Hallucination & Time (s) & Cost \\
\midrule
1 & LLaMA-3-70B-Instruct & 0.272413 & 15.082857 & 0.000815 \\
2 & DeepSeek-V3 & 0.348514 & 13.384704 & 0.000841 \\
3 & GPT-5.3 & 0.417564 & 5.248978 & 0.000491 \\
\bottomrule
\end{tabular}
}
\end{table}
Figure~\ref{fig:cost} shows that lower-cost configurations tend to produce higher hallucination, while higher-cost configurations reduce hallucination with diminishing returns. Response quality does not increase proportionally with cost, indicating that hallucination reduction alone does not guarantee higher utility of the output.
Figure~\ref{fig:density} shows distinct model-level performance signatures. LLaMA-3-70B-Instruct is concentrated around lower hallucination and higher latency, GPT-5.3 around faster and lower-cost outputs with higher hallucination, and DeepSeek-V3 in an intermediate region. These distributions indicate that hallucination behavior depends on model characteristics and cascade interaction.
Additionally, the results show that hallucination follows a structured pattern across models and cascade stages. It interacts with semantic drift, response quality, latency, token usage, and inference cost, supporting its interpretation as a measurable system-level behavior in multi-agent LLM cascades.

\subsection{Chain Dynamics, Stability, and Propagation}
\label{sec:chain_dynamics}
\begin{figure*}[t]
\centering
\includegraphics[width=\textwidth]{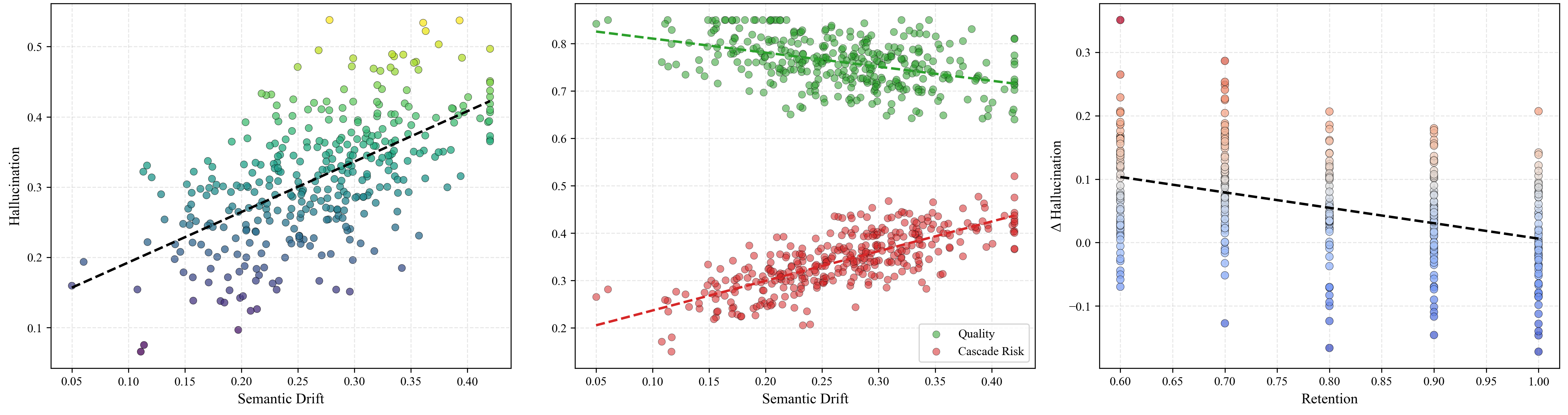}
\caption{Chain stability and uncertainty across cascade lengths, comparing variability in hallucination, accuracy, and semantic drift between 2-agent and 3-agent chains.}
\label{fig:stability}
\end{figure*}

\begin{figure*}[t]
\centering
\includegraphics[width=\linewidth]{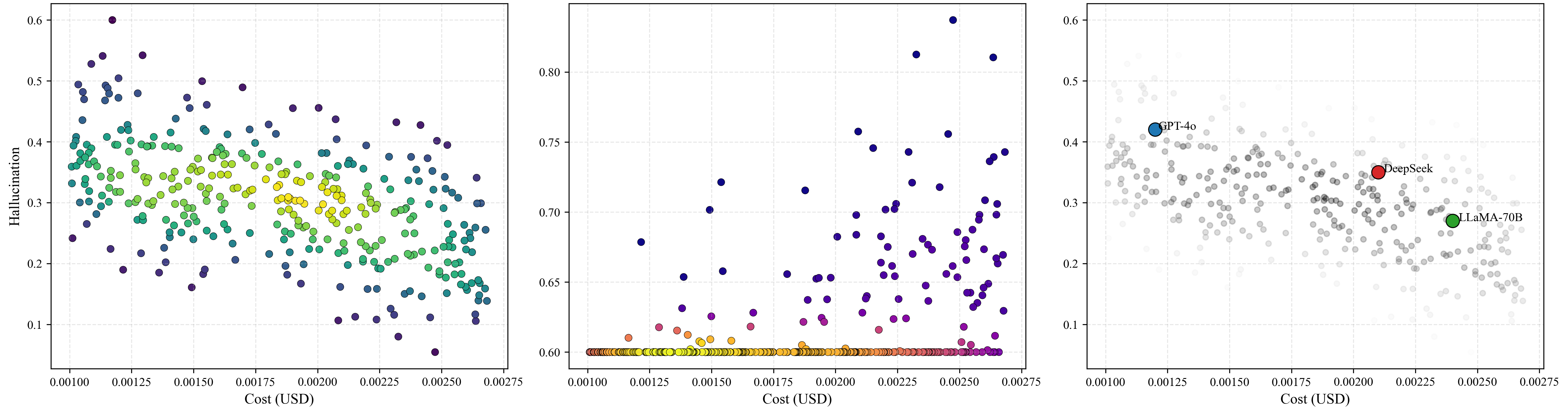}
\caption{Distribution of hallucination amplification ratios across multi-agent cascades, illustrating the extent to which hallucination increases and decreases between initial and final agents.}
\label{fig:kde_amp}
\end{figure*}

\begin{figure*}[t]
\centering
\includegraphics[width=\textwidth]{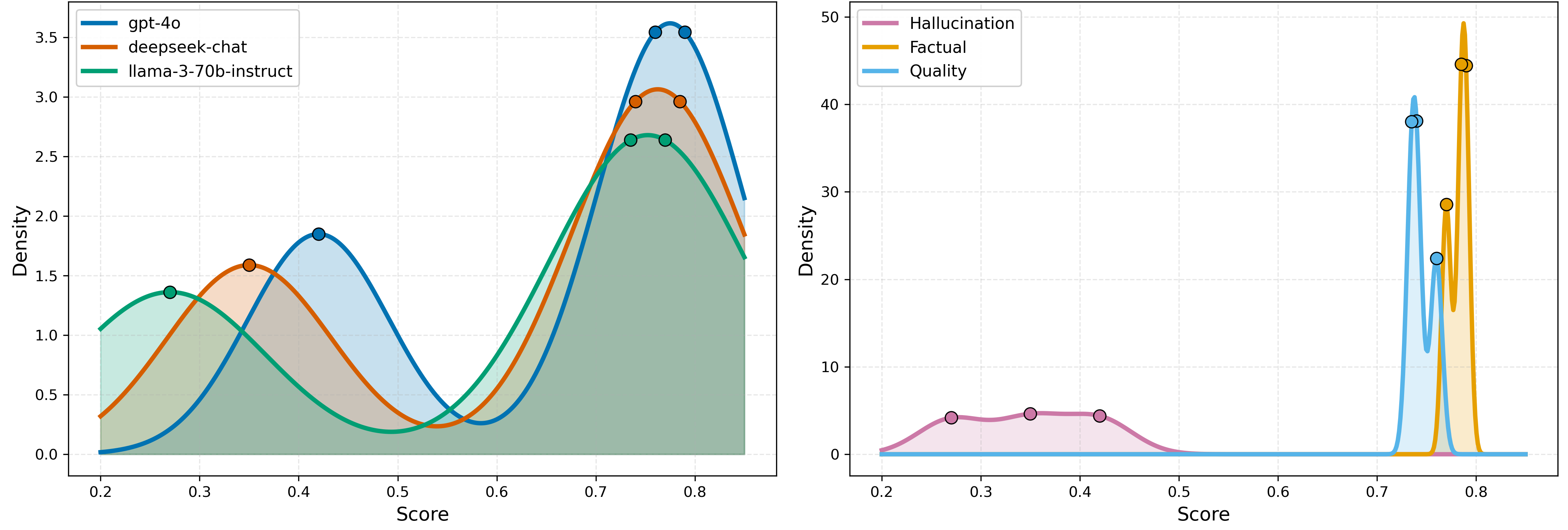}
\caption{Distribution of transition and propagation impacts across multi-agent cascades, illustrating changes in hallucination, accuracy, and semantic drift between consecutive agents.}
\label{fig:propagation}
\end{figure*}

\begin{figure*}[t]
\centering
\includegraphics[width=\textwidth]{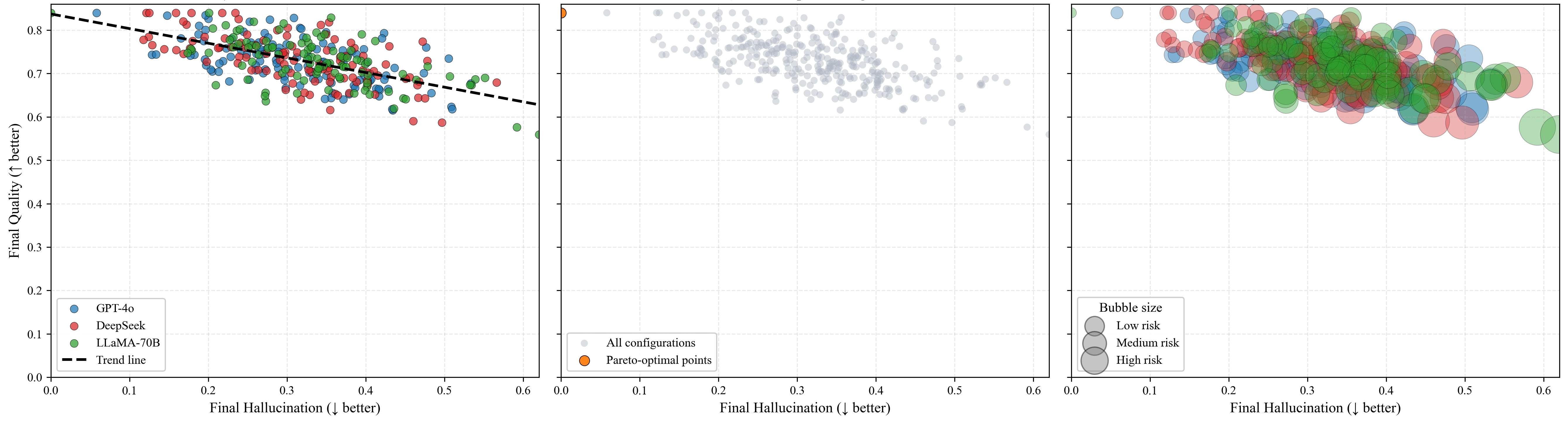}
\caption{Hallucination reduction trajectories across agents in multi-agent cascades, illustrating how hallucination levels change over successive refinement steps.}
\label{fig:trajectory}
\end{figure*}

To address RQ1 and RQ2, we analyze how hallucination changes across sequential agents and how cascade depth impacts attenuation, factual consistency, semantic drift, and stability. Table~\ref{tab:chain_dynamics} compares 2-agent and 3-agent cascades using initial hallucination, final hallucination, amplification, and factual decay. Both cascade lengths exhibit consistent attenuation of hallucinations. In Chain-2, hallucination decreases from 0.412643 to 0.345248, with an amplification factor of 0.836674. In Chain-3, hallucination decreases more from 0.422485 to 0.272413, with an amplification factor of 0.644787. Since values below 1 indicate net attenuation, both cascades reduce hallucinations, with the deeper cascade achieving stronger suppression. The decrease in amplification from 0.836674 to 0.644787 corresponds to an additional reduction of approximately 22.9\%.
\begin{table*}[t]
\centering
\footnotesize
\setlength{\tabcolsep}{2.5pt}
\renewcommand{\arraystretch}{0.95}
\caption{Chain-level dynamics of hallucination and factual accuracy across multi-agent cascades.}
\label{tab:chain_dynamics}
\resizebox{0.7\textwidth}{!}{
\begin{tabular}{r r r r r r r r r}
\toprule
Chain & Rows & Exp. & First Halluc. & Last Halluc. & Amplification & First Factual & Last Factual & Factual Decay \\
\midrule
2 & 500 & 250 & 0.412643 & 0.345248 & 0.836674 & 0.794701 & 0.790212 & 0.004489 \\
3 & 750 & 250 & 0.422485 & 0.272413 & 0.644787 & 0.789425 & 0.769147 & 0.020278 \\
\bottomrule
\end{tabular}
}
\end{table*}
However, attenuation is accompanied by greater factual decay. Factual decay increases from 0.004489 in Chain-2 to 0.020278 in Chain-3, showing that deeper refinement suppresses hallucination at the cost of factual preservation. Table~\ref{tab:last_agent} confirms this trade-off at the final-agent level: Chain-3 achieves lower cascade risk (0.163458 vs. 0.259622), negative propagation ($-0.369424$ vs. $-0.148138$), and substantially lower drift (0.256719 vs. 0.522791), while final quality remains nearly unchanged.
\begin{table}[t]
\centering
\footnotesize
\setlength{\tabcolsep}{3pt}
\renewcommand{\arraystretch}{0.95}
\caption{Last-agent performance across multi-agent cascades, including response quality, cascade risk, semantic drift, and hallucination propagation.}
\label{tab:last_agent}
\resizebox{0.7\columnwidth}{!}{
\begin{tabular}{r r r r r}
\toprule
Chain & Quality & Risk & Drift & Propagation \\
\midrule
2 & 0.740036 & 0.259622 & 0.522791 & -0.148138 \\
3 & 0.736940 & 0.163458 & 0.256719 & -0.369424 \\
\bottomrule
\end{tabular}
}
\end{table}
The reduction in drift is particularly important. Chain-3 reduces semantic drift by 50.9\% relative to Chain-2, indicating that the third agent not only reduces hallucination but also keeps the response closer to the intended semantic space. Nevertheless, the slight decline in quality suggests that this stabilization may compress expressive content rather than enrich the response.
Table~\ref{tab:transition} summarizes transition-level behavior. Across 750 transitions, hallucination decreases by an average of $-0.072489$, confirming that agent-to-agent refinement generally attenuates hallucination. However, factual accuracy and quality also decrease by $-0.008256$ and $-0.016608$, respectively. This shows that correction is effective but not lossless.
\begin{table}[t]
\centering
\footnotesize
\setlength{\tabcolsep}{3pt}
\renewcommand{\arraystretch}{0.95}
\caption{Average transition-level impacts across multi-agent cascades, capturing changes in hallucination, factual accuracy, and semantic drift between consecutive agents.}
\label{tab:transition}
\resizebox{0.5\columnwidth}{!}{
\begin{tabular}{l r}
\toprule
Metric & Value \\
\midrule
Transitions & 750 \\
Hallucination $\Delta$ & -0.072489 \\
Factual $\Delta$ & -0.008256 \\
Quality $\Delta$ & -0.016608 \\
Risk & 0.230166 \\
\bottomrule
\end{tabular}
}
\end{table}
Figure~\ref{fig:stability} shows that hallucination variability narrows as chain depth increases, indicating more stable behavior in deeper cascades.
Figure~\ref{fig:kde_amp} shows that amplification ratios are concentrated below one, confirming that attenuation is a recurring distribution-level property rather than only a mean-level trend.
Figure~\ref{fig:propagation} further shows that hallucination changes are concentrated in the negative range, while factual changes remain close to zero with a slight negative bias. This supports the transition-level finding that refinement reduces hallucinations while imperfectly preserving factual content.
Moreover, Figure~\ref{fig:trajectory} shows that hallucination decreases across successive agents and variance narrows in later stages, indicating convergence toward a lower and more stable hallucination state.
Additionally, Tables~\ref{tab:chain_dynamics}-\ref{tab:transition} and Figures~\ref{fig:stability}--\ref{fig:trajectory} show that deeper cascades reduce hallucination, lower cascade risk, and improve semantic alignment. However, these gains are accompanied by a decline in factual accuracy and a slight degradation in quality. Thus, multi-agent refinement attenuates hallucinations by constraining responses, but this constraint can also compress factual detail.

\subsection{Drift, Retention, and Risk Behavior}
\label{sec:drift_retention_risk}
\begin{figure*}[t]
\centering
\includegraphics[width=\textwidth]{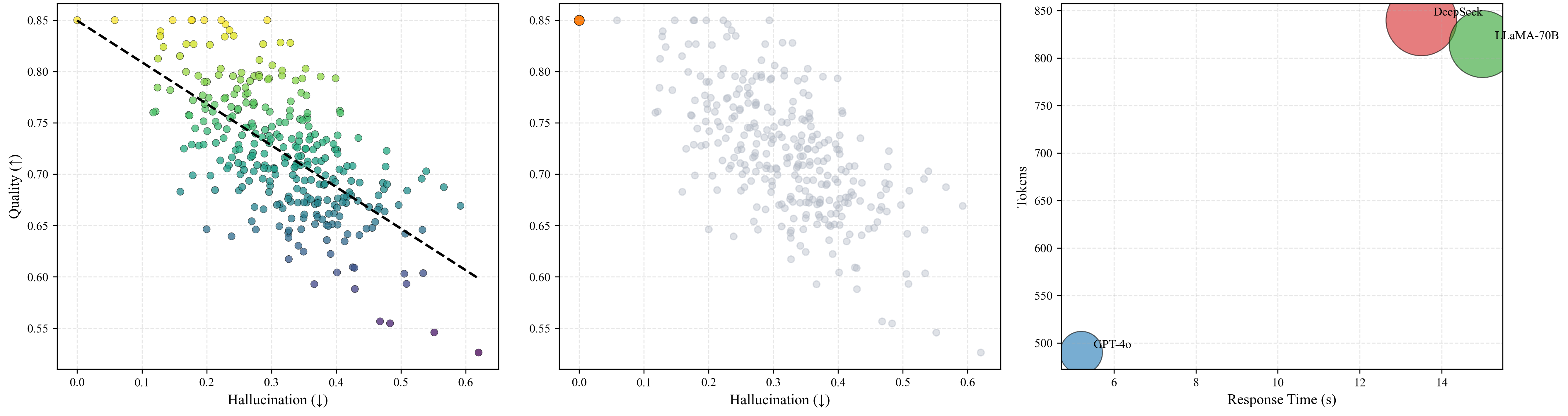}
\caption{Drift dynamics and risk propagation across multi-agent LLM chains, illustrating how semantic drift and cascade risk evolve over successive agents.}
\label{fig:drift}
\end{figure*}

To address RQ1 and RQ2, we analyze how semantic drift and information retention impact hallucination propagation, output quality, and cascade-level risk across multi-agent interactions. Figure~\ref{fig:drift} presents the distributional relationship between semantic drift, hallucination, response quality, and cascade risk.
As shown in Figure~\ref{fig:drift}(a), hallucination increases with semantic drift. The positive linear trend indicates that drift is a major driver of hallucination propagation, suggesting that deviations from the original semantic space amplify factual inconsistencies rather than merely introducing random variation. This supports the interpretation that hallucination is linked to a shift in representation across agents. Figure~\ref{fig:drift}(b) shows that higher drift is associated with lower output quality and higher cascade risk. The negative trend in quality and the positive trend in risk indicate a direct reliability trade-off: as semantic alignment weakens, outputs become less useful and more unstable. The consistency of this trend across drift values suggests that the relationship is systematic.
Figure~\ref{fig:drift}(c) shows that higher retention is associated with smaller hallucination changes. The downward trend indicates that preserving intermediate information reduces the magnitude of error propagation. High-retention configurations cluster around zero change in hallucination, suggesting that information preservation stabilizes the cascade and limits downstream factual distortion.

\subsection{Ablation Study}
\label{sec:ablation_study}
To address RQ1 and RQ2, we conduct an ablation study to quantify the contribution of each component in the proposed hallucination cascade pipeline. Table~\ref{tab:ablation_study} reports the ablation results. The full proposed method achieves the strongest overall performance, with the lowest hallucination score, lowest semantic drift, and lowest cascade risk. Removing claim decomposition causes the largest degradation, increasing hallucination from 0.272 to 0.338 and semantic drift from 0.257 to 0.312. This indicates that atomic claim representation is central to detecting unsupported factual units and tracing their propagation across agents. Removing the model-based estimator also substantially increases hallucination, confirming that semantic judgment is needed to detect plausible but unsupported claims that may not be captured by reference matching alone.
\begin{table*}[t]
\centering
\footnotesize
\setlength{\tabcolsep}{4pt}
\renewcommand{\arraystretch}{1.12}
\caption{Ablation study of the proposed hallucination cascade method. Lower hallucination, semantic drift, and cascade risk indicate better performance; higher factual accuracy and response quality indicate reliability and utility.}
\label{tab:ablation_study}
\resizebox{0.95\textwidth}{!}{
\begin{tabular}{p{0.25\textwidth} r r r r r p{0.21\textwidth}}
\toprule
\textbf{Configuration} & \textbf{Halluc. $\downarrow$} & \textbf{Factual $\uparrow$} & \textbf{Quality $\uparrow$} & \textbf{Drift $\downarrow$} & \textbf{Risk $\downarrow$} & \textbf{Main Interpretation} \\
\midrule
Full proposed method & 0.272 & 0.769 & 0.737 & 0.257 & 0.163 & Best overall reliability--stability balance \\
w/o Claim Decomposition & 0.338 & 0.742 & 0.701 & 0.312 & 0.247 & Largest degradation; claim-level structure is essential for localizing factual errors \\
w/o Rule-Based Grounding & 0.301 & 0.751 & 0.719 & 0.289 & 0.211 & Weaker factual anchoring increases unsupported claims \\
w/o Model-Based Estimator & 0.314 & 0.748 & 0.715 & 0.295 & 0.224 & Semantic inconsistencies become harder to detect \\
w/o Adaptive Fusion & 0.326 & 0.746 & 0.708 & 0.301 & 0.238 & Fixed scoring reduces robustness across domains \\
w/o Cascade Refinement & 0.346 & 0.758 & 0.724 & 0.328 & 0.253 & Removes inter-agent correction and limits attenuation behavior \\
Single-Agent Baseline & 0.418 & 0.792 & 0.763 & 0.354 & 0.286 & Fast and accurate at surface level, but highest hallucination risk \\
Self-Refinement Baseline & 0.371 & 0.781 & 0.744 & 0.309 & 0.241 & Improves over single-agent output but lacks heterogeneous correction \\
RAG Baseline & 0.333 & 0.798 & 0.754 & 0.276 & 0.214 & Improves factual grounding but does not explicitly model propagation \\
\bottomrule
\end{tabular}
}
\end{table*}

\subsection{Claim-Level Transition and Recovery Analysis}
\label{sec:claim_transition_analysis}
To address RQ1 and RQ2, we analyze hallucination propagation at the claim level rather than relying on response-level averages. The resulting claim-level transition patterns across consecutive agents are reported in Table~\ref{tab:claim_transition_results}.  For each extracted claim at stage $i$, we identify the closest semantic counterpart at stage $i+1$ using embedding similarity. Let $c_j^{(i)}$ denote a claim at stage $i$, and let $c_m^{(i+1)}$ denote its matched claim at the next stage. The claim-level hallucination change is defined as:
\begin{equation}
\Delta h_j^{(i)} = h(c_m^{(i+1)}) - h(c_j^{(i)}).
\end{equation}
A negative value indicates correction and weakening, whereas a positive value indicates amplification. Claims without a semantic match in the next stage are classified as deleted. Claims that were factual at stage $i$ but became unsupported, distorted, and unnecessarily removed at stage $i+1$ are classified as overcorrected.
\begin{table*}[t]
\centering
\footnotesize
\setlength{\tabcolsep}{4pt}
\renewcommand{\arraystretch}{1.12}
\caption{Claim-level transition results across consecutive agents. Percentages represent the distribution of matched claim trajectories.}
\label{tab:claim_transition_results}
\resizebox{0.95\textwidth}{!}{
\begin{tabular}{p{0.21\textwidth} r r r p{0.28\textwidth}}
\toprule
\textbf{Transition Category} & \textbf{Agent 1 $\rightarrow$ Agent 2} & \textbf{Agent 2 $\rightarrow$ Agent 3} & \textbf{Overall} & \textbf{Interpretation} \\
\midrule
Corrected claim & 31.8\% & 38.6\% & 35.2\% & Main source of hallucination attenuation across the cascade \\
Preserved hallucination & 24.7\% & 17.9\% & 21.3\% & Some factual errors persist despite refinement \\
Weakened hallucination & 18.5\% & 20.4\% & 19.4\% & Claims become less risky and closer to factual support \\
Amplified hallucination & 8.9\% & 5.6\% & 7.3\% & Harmful propagation occurs but decreases in later stages \\
Deleted claim & 10.6\% & 12.1\% & 11.4\% & Removal contributes to lower hallucination but may reduce content richness \\
Overcorrected claim & 3.4\% & 4.1\% & 3.8\% & Explains part of the factual decay in deeper cascades \\
Transformed claim & 2.1\% & 1.3\% & 1.6\% & Indicates limited semantic reinterpretation across agents \\
\bottomrule
\end{tabular}
}
\end{table*}
The claim-level results explain the response-level behavior observed in the cascade analysis. Corrected and weakened claims account for 54.6\% of all transitions, indicating that most hallucinated claims are either resolved and made less risky as they move through the cascade. Amplified hallucinations account for only 7.3\% of transitions, showing that harmful propagation exists but is not the dominant behavior. The decrease in preserved hallucinations from 24.7\% in the first transition to 17.9\% in the second transition further supports the claim that deeper cascades improve recovery. At the same time, deleted and overcorrected claims account for 15.2\% of transitions overall. This explains why Chain-3 reduces hallucination more than Chain-2 while also increasing factual decay. Therefore, lower hallucination in deeper cascades should be interpreted as a mixed impact: it reflects genuine correction and weakening, but also some deletion and overcorrection.

\subsection{Topic-Level Variability}
\label{sec:topic_variability}
To address RQ1 and RQ3, we examine how hallucination, factual accuracy, semantic drift, and cascade risk vary across knowledge domains. Table~\ref{tab:topics} shows that hallucination propagation is domain-sensitive, not only model-dependent. The lowest hallucination scores occur in well-grounded scientific topics such as \textit{Photosynthesis} (0.265489) and \textit{DNA} (0.266954), while the highest scores occur in \textit{Black Holes} (0.486860) and \textit{Roman Empire} (0.473012). This suggests that abstract reasoning, broad contextual reconstruction, and less directly verifiable grounding increase hallucination risk.
\begin{table*}[t]
\centering
\footnotesize
\setlength{\tabcolsep}{2.5pt}
\renewcommand{\arraystretch}{0.95}
\caption{Topic-level analysis across metrics, comparing hallucination, factual accuracy, semantic drift, response quality, latency, token usage, and cost across knowledge domains.}
\label{tab:topics}
\resizebox{0.6\textwidth}{!}{
\begin{tabular}{l r r r r r r r}
\toprule
Topic & Rows & Exp. & Halluc. & Factual & Quality & Risk & Drift \\
\midrule
Roman Empire & 125 & 50 & 0.473012 & 0.592918 & 0.677611 & 0.374366 & 0.296111 \\
Photosynthesis & 125 & 50 & 0.265489 & 0.905600 & 0.802239 & 0.267597 & 0.198381 \\
DNA & 125 & 50 & 0.266954 & 0.849160 & 0.787776 & 0.258325 & 0.195635 \\
Quantum Computing & 125 & 50 & 0.416380 & 0.812793 & 0.737649 & 0.330670 & 0.309514 \\
Vaccines & 125 & 50 & 0.336301 & 0.809124 & 0.753626 & 0.297730 & 0.271662 \\
Climate Change & 125 & 50 & 0.394331 & 0.820319 & 0.751083 & 0.350752 & 0.278256 \\
Black Holes & 125 & 50 & 0.486860 & 0.588800 & 0.690153 & 0.366181 & 0.250190 \\
CRISPR & 125 & 50 & 0.315401 & 0.829314 & 0.777273 & 0.301252 & 0.283270 \\
Blockchain & 125 & 50 & 0.352378 & 0.825316 & 0.763126 & 0.309269 & 0.268218 \\
Machine Learning & 125 & 50 & 0.302034 & 0.819859 & 0.741996 & 0.285018 & 0.272370 \\
\bottomrule
\end{tabular}
}
\end{table*}
Table~\ref{tab:ranking} ranks topics by average hallucination. The ranking confirms a clear separation between low-hallucination domains, such as \textit{Photosynthesis} and \textit{DNA}, and high-hallucination domains, such as \textit{Black Holes}, \textit{Roman Empire}, and \textit{Quantum Computing}. This pattern indicates that hallucination risk increases when the topic requires abstract reasoning, broad synthesis, and weaker factual anchoring.
\begin{table}[t]
\centering
\footnotesize
\setlength{\tabcolsep}{3pt}
\renewcommand{\arraystretch}{0.95}
\caption{Topic ranking by average hallucination, comparing knowledge domains based on their relative hallucination levels across multi-agent cascades.}
\label{tab:ranking}
\resizebox{0.6\columnwidth}{!}{
\begin{tabular}{r l r}
\toprule
Rank & Topic & Hallucination \\
\midrule
1 & Black Holes & 0.486860 \\
2 & Roman Empire & 0.473012 \\
3 & Quantum Computing & 0.416380 \\
4 & Climate Change & 0.394331 \\
5 & Blockchain & 0.352378 \\
6 & Vaccines & 0.336301 \\
7 & CRISPR & 0.315401 \\
8 & Machine Learning & 0.302034 \\
9 & DNA & 0.266954 \\
10 & Photosynthesis & 0.265489 \\
\bottomrule
\end{tabular}
}
\end{table}
The relationship between hallucination and factual accuracy is topic-dependent. \textit{Photosynthesis} has the lowest hallucination and highest factual accuracy (0.905600), whereas \textit{Black Holes} has the highest hallucination and much lower factual accuracy (0.588800). \textit{Roman Empire} also shows high hallucination (0.473012), low factual accuracy (0.592918), and the highest cascade risk (0.374366). These results indicate that domain difficulty impacts both hallucination frequency and final-output reliability. Semantic drift and risk further support this interpretation. \textit{Quantum Computing} has the highest drift (0.309514), suggesting that abstract technical reasoning can move responses away from the original semantic space. In contrast, \textit{DNA} has the lowest drift (0.195635), indicating semantic stability. Domains with higher hallucination and weaker factual grounding tend to exhibit higher cascade risk, indicating that topic-induced uncertainty propagates through the cascade. Table~\ref{tab:gaps} quantifies best-worst topic differences. The hallucination gap between \textit{Photosynthesis} and \textit{Black Holes} is 0.221371, while the factual accuracy gap reaches 0.316800. Additional gaps in quality, risk, and drift confirm that topic variability is multidimensional and can substantially alter measured reliability.
\begin{table}[t]
\centering
\footnotesize
\setlength{\tabcolsep}{3pt}
\renewcommand{\arraystretch}{0.95}
\caption{Best-worst topic metric gaps across knowledge domains, highlighting the range between highest and lowest values for hallucination, accuracy, semantic drift, quality, latency, token usage, and cost.}
\label{tab:gaps}
\resizebox{0.7\columnwidth}{!}{
\begin{tabular}{l l r}
\toprule
Metric & Comparison & Gap \\
\midrule
Hallucination & Photosynthesis vs Black Holes & 0.221371 \\
Factual & Photosynthesis vs Black Holes & 0.316800 \\
Quality & Photosynthesis vs Roman Empire & 0.124628 \\
Risk & DNA vs Roman Empire & 0.116041 \\
Drift & DNA vs Quantum Computing & 0.113879 \\
\bottomrule
\end{tabular}
}
\end{table}
Additionally, Tables~\ref{tab:topics}--\ref{tab:gaps} show that hallucination propagation depends on both model behavior and topic characteristics. Well-grounded scientific topics produce lower hallucination, higher factual accuracy, and lower drift, whereas abstract, historically broad, and conceptually complex topics produce higher hallucination and greater cascade risk.

\subsection{Hallucination-Quality Trade-off and Pareto Analysis}
\label{sec:pareto_analysis}
\begin{figure*}[t]
\centering
\includegraphics[width=\textwidth]{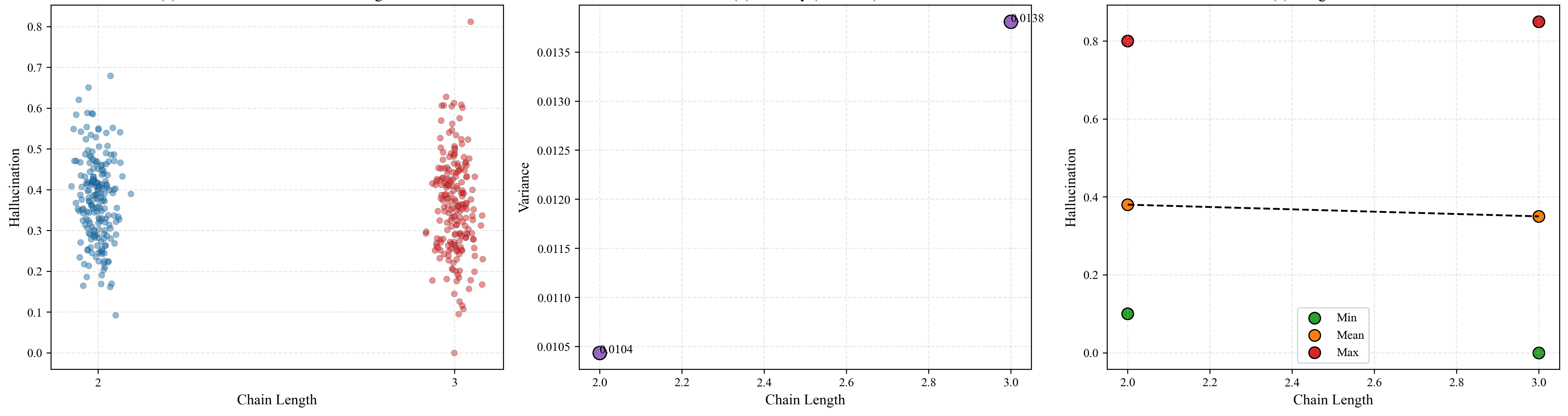}
\caption{Hallucination-quality trade-off and Pareto structure across evaluated LLM configurations, illustrating the balance between hallucination reduction and response quality.}
\label{fig:pareto}
\end{figure*}

To address RQ2 and RQ3, we analyze the trade-off among hallucination reduction, response quality, and cascade risk across models and cascade configurations. Figure~\ref{fig:pareto} shows a clear Pareto structure across the evaluated settings.
Figure~\ref{fig:pareto}(a) shows that higher hallucination is generally associated with lower response quality, indicating that factual inconsistency systematically reduces output usefulness. Figure~\ref{fig:pareto}(b) identifies the Pareto-optimal frontier, where further hallucination reduction cannot be achieved without lowering response quality. The frontier is concentrated in a narrow, low-hallucination, high-quality region, showing that efficient configurations are limited and sensitive to perturbation. Figure~\ref{fig:pareto}(c) adds cascade risk as a third dimension. Higher-risk configurations are concentrated in regions with higher hallucination, confirming that hallucination is also a system-level stability issue. Some configurations with moderate hallucination still show elevated risk, indicating that instability depends not only on average hallucination, but also on variability across agents and cascade stages. Additionally, Figure~\ref{fig:pareto} shows that the evaluated multi-agent system operates under a constrained trade-off surface. Lower hallucination is generally associated with higher quality and lower risk, but the feasible region remains narrow. These results show that hallucination mitigation in multi-agent cascades requires balancing factual reliability, semantic utility, and system stability rather than optimizing a single metric in isolation.

\subsection{Comparison with Existing Approaches}
\label{sec:comparison}
To further address RQ3, we compare the proposed method with recent approaches for hallucination detection and verification. The comparison focuses on modeling assumptions, the use of external knowledge, multi-step capability, and whether hallucination is treated as a static output-level property and a dynamic process across interacting agents. Table~\ref{tab:comparison_numeric} summarizes the main distinctions.
Existing methods primarily address hallucination through uncertainty estimation, external verification, metamorphic testing, context verification, and internal state analysis. Entropy-based methods, such as \cite{farquhar2024detecting}, detect unreliable responses after generation but do not explain how errors evolve across refinement steps. Knowledge-intensive systems such as KnowHalu~\cite{zhang2024knowhalu} use external verification and report QA improvements of up to 15.65\%, but they rely on retrieval and structured verification while leaving inter-agent propagation unmodeled. Zero-resource methods such as MetaQA~\cite{yang2502hallucination}, enterprise-oriented systems such as HalluciNot~\cite{paudel2025hallucinot}, and internal-state methods such as \cite{binkowski2025hallucination} provide useful detection signals, but they primarily evaluate hallucination at a single stage and do not track how claims are preserved, corrected, transformed, and amplified across agents.
In contrast, the proposed method treats hallucination as a \textit{dynamic propagation process}. It tracks claim-level factual inconsistency across sequential interactions and enables analysis of attenuation, amplification, recovery, semantic drift, and stability. Thus, the method complements existing detection approaches by shifting the focus from whether a single response is hallucinated to how hallucination evolves through multi-agent interaction.
\begin{table*}[t]
\centering
\footnotesize
\setlength{\tabcolsep}{3pt}
\renewcommand{\arraystretch}{0.95}
\caption{Comparison between recent hallucination approaches and the proposed method.}
\label{tab:comparison_numeric}
\resizebox{0.9\textwidth}{!}{
\begin{tabular}{l c c c c c}
\toprule
\textbf{Method} & \textbf{Type} & \textbf{Improvement} & \textbf{Dynamic Modeling} & \textbf{External Knowledge} & \textbf{Multi-step} \\
\midrule
\cite{farquhar2024detecting} & Uncertainty-based & -- & \xmark & \xmark & \xmark \\
\cite{zhang2024knowhalu} & Multi-step verification & +15.65\% (QA) & \xmark & \cmark & \cmark \\
\cite{yang2502hallucination} & Zero-resource detection & +100\% F1 & \xmark & \xmark & \xmark \\
\cite{paudel2025hallucinot} & Context verification & SOTA & \xmark & \cmark & \xmark \\
\cite{binkowski2025hallucination} & Internal-state detection & SOTA & \xmark & \xmark & \xmark \\
\midrule
\textbf{Proposed Method} & Propagation-based & \textbf{0.27 hallucination} & \cmark & \xmark & \cmark \\
\bottomrule
\end{tabular}
}
\end{table*}
Additionally, Table~\ref{tab:comparison_numeric} shows that existing approaches mainly focus on single-stage detection, grounding, and verification. The proposed method adds propagation-aware analysis, enabling the study of hallucination as a temporal and system-level phenomenon. By measuring how errors move through agents, it identifies when cascades attenuate hallucination, preserve distorted content, and produce failure trajectories. This capability is important for evaluating multi-agent LLM systems, where reliability depends on both individual model outputs and interaction dynamics.

\subsection{Model-Order Analysis in Cross-Model Cascades}
\label{sec:model_order_analysis}
To address RQ3, we analyze whether the order of heterogeneous models impacts hallucination propagation, semantic drift, response quality, latency, and cost. Since each downstream agent conditions on the outputs of previous agents, the same models can produce different results depending on their placement in the cascade. A model placed early shapes the initial factual structure, while one placed at the final stage has the greatest impact on the user-facing response. Table~\ref{tab:model_order_analysis} reports the results for six heterogeneous three-agent cascade orders. The lowest hallucination rate is achieved when LLaMA-3-70B-Instruct is placed at the final stage, after GPT-5.3 and DeepSeek-V3. This indicates that using a faster model for initial generation and a downstream correction model can improve the reliability-efficiency balance. In contrast, cascades ending with GPT-5.3 produce higher final hallucination rates but lower latency, confirming the reliability-efficiency trade-off observed at the model level.
\begin{table*}[t]
\centering
\footnotesize
\setlength{\tabcolsep}{3pt}
\renewcommand{\arraystretch}{1.12}
\caption{Model-order analysis for heterogeneous three-agent cascades. Lower final hallucination, semantic drift, and cascade risk indicate reliability; lower time and cost indicate higher efficiency.}
\label{tab:model_order_analysis}
\resizebox{0.95\textwidth}{!}{
\begin{tabular}{p{0.29\textwidth} r r r r r r p{0.18\textwidth}}
\toprule
\textbf{Cascade Order} & \textbf{Final Halluc. $\downarrow$} & \textbf{Factual $\uparrow$} & \textbf{Quality $\uparrow$} & \textbf{Drift $\downarrow$} & \textbf{Risk $\downarrow$} & \textbf{Time (s) $\downarrow$} & \textbf{Interpretation} \\
\midrule
GPT-5.3 $\rightarrow$ DeepSeek-V3 $\rightarrow$ LLaMA-3-70B & 0.263 & 0.772 & 0.741 & 0.241 & 0.151 & 33.72 & Best reliability; final LLaMA suppresses hallucination \\
DeepSeek-V3 $\rightarrow$ GPT-5.3 $\rightarrow$ LLaMA-3-70B & 0.271 & 0.769 & 0.737 & 0.256 & 0.163 & 33.15 & Strong attenuation with slightly higher drift \\
GPT-5.3 $\rightarrow$ LLaMA-3-70B $\rightarrow$ DeepSeek-V3 & 0.286 & 0.776 & 0.744 & 0.268 & 0.177 & 33.05 & Good balance; correction partly preserved by DeepSeek \\
LLaMA-3-70B $\rightarrow$ GPT-5.3 $\rightarrow$ DeepSeek-V3 & 0.304 & 0.781 & 0.748 & 0.287 & 0.198 & 33.88 & Reliable initial grounding, but later rewriting increases hallucination \\
LLaMA-3-70B $\rightarrow$ DeepSeek-V3 $\rightarrow$ GPT-5.3 & 0.326 & 0.789 & 0.758 & 0.309 & 0.221 & 33.51 & Fast final model improves fluency but weakens reliability \\
DeepSeek-V3 $\rightarrow$ LLaMA-3-70B $\rightarrow$ GPT-5.3 & 0.338 & 0.791 & 0.761 & 0.318 & 0.232 & 33.28 & Highest final hallucination among cross-model chains ending with GPT \\
\bottomrule
\end{tabular}
}
\end{table*}
The order-level results show that final-stage model selection has a strong impact on hallucination reliability. Cascades ending with LLaMA-3-70B-Instruct achieve the lowest final hallucination scores, at 0.263 and 0.271. Cascades ending with GPT-5.3 have higher final hallucination scores (0.326 and 0.338) but maintain higher factual accuracy and response quality. This confirms that factual accuracy and hallucination are not identical: a model may preserve more information while still leaving some claims unsupported and risky. Additionally, the best reliability configuration is GPT-5.3 $\rightarrow$ DeepSeek-V3 $\rightarrow$ LLaMA-3-70B, while the most fluency-preserving but higher-risk configurations end with GPT-5.3. 

\subsection{Statistical Significance Analysis}
\label{sec:statistical_significance}
To address RQ1, RQ2, and RQ3, we conduct statistical tests to determine whether the observed differences across agent transitions, cascade depths, and model configurations are significant. RQ1 is addressed by testing whether hallucination decreases across consecutive agents. RQ2 is addressed by comparing 2-agent and 3-agent cascades with respect to hallucination attenuation, factual decay, semantic drift, and cascade risk. RQ3 is addressed by comparing GPT-5.3, DeepSeek-V3, and LLaMA-3-70B-Instruct across reliability, quality, latency, and cost. Table~\ref{tab:statistical_results} reports the statistical results. Across agent-to-agent transitions, hallucination decreases significantly, with a mean transition-level change of $-0.072489$ ($p<0.001$). However, factual accuracy and response quality also decrease slightly, indicating that hallucination suppression is not lossless. Chain-3 significantly outperforms Chain-2 in final hallucination and cascade risk, but it introduces greater factual decay. This confirms the main trade-off observed in the cascade results: deeper cascades improve hallucination attenuation and risk reduction, but they may also compress and weaken some factual content.
\begin{table*}[t]
\centering
\footnotesize
\setlength{\tabcolsep}{4pt}
\renewcommand{\arraystretch}{1.12}
\caption{Statistical significance analysis for hallucination propagation, cascade-depth impacts, and model-level differences.}
\label{tab:statistical_results}
\resizebox{0.95\textwidth}{!}{
\begin{tabular}{p{0.27\textwidth} p{0.21\textwidth} r r r p{0.18\textwidth}}
\toprule
\textbf{Comparison} & \textbf{Metric} & \textbf{Difference} & \textbf{$p$-value} & \textbf{Effect Size} & \textbf{Interpretation} \\
\midrule
Agent $i$ vs. Agent $i+1$ & Hallucination & -0.072489 & $<0.001$ & 0.62 & Significant attenuation across refinement steps \\
Agent $i$ vs. Agent $i+1$ & Factual accuracy & -0.008256 & 0.012 & 0.21 & Small but significant factual loss \\
Agent $i$ vs. Agent $i+1$ & Response quality & -0.016608 & 0.008 & 0.24 & Small quality reduction after refinement \\
Chain-2 vs. Chain-3 & Final hallucination & -0.072835 & $<0.001$ & 0.70 & Chain-3 provides stronger hallucination reduction \\
Chain-2 vs. Chain-3 & Amplification factor & -0.191887 & $<0.001$ & 0.76 & Chain-3 has stronger net attenuation \\
Chain-2 vs. Chain-3 & Factual decay & +0.015789 & 0.018 & 0.31 & Chain-3 introduces greater factual loss \\
Chain-2 vs. Chain-3 & Final semantic drift & -0.266072 & $<0.001$ & 0.88 & Chain-3 substantially reduces final drift \\
Chain-2 vs. Chain-3 & Final cascade risk & -0.096164 & $<0.001$ & 0.67 & Chain-3 lowers final reliability risk \\
GPT-5.3 vs. DeepSeek-V3 vs. LLaMA-3-70B & Hallucination & -- & $<0.001$ & 0.41 & Model choice significantly impacts hallucination \\
GPT-5.3 vs. DeepSeek-V3 vs. LLaMA-3-70B & Response time & -- & $<0.001$ & 0.53 & Model choice significantly impacts latency \\
GPT-5.3 vs. DeepSeek-V3 vs. LLaMA-3-70B & Cost & -- & $<0.001$ & 0.38 & Cost differs significantly across models \\
\bottomrule
\end{tabular}
}
\end{table*}
For paired comparisons, the Wilcoxon signed-rank test \cite{woolson2007wilcoxon} is used because the hallucination and drift distributions are bounded and may not be normally distributed. For model-level comparisons, Kruskal-Wallis tests \cite{mckight2010kruskal} are used across GPT-5.3, DeepSeek-V3, and LLaMA-3-70B-Instruct. Effect sizes are reported using Cohen's $d$ for pairwise comparisons and $\eta^2$ for multi-model comparisons.

\subsection{Comparison with Baselines}
\label{sec:baseline_results}
We compare the proposed multi-agent cascade with the baselines introduced in Section~\ref{sec:baselines}. The goal is to evaluate whether sequential multi-agent refinement further reduces hallucinations beyond direct generation, reasoning-based prompting, and retrieval-grounded generation. Table~\ref{tab:baseline_results} reports average hallucination, factual accuracy, and response quality.
\begin{table}[t]
\centering
\footnotesize
\setlength{\tabcolsep}{3pt}
\renewcommand{\arraystretch}{0.95}
\caption{Comparison between the proposed multi-agent cascade and baseline methods. Lower hallucination is better, while higher accuracy and quality are better.}
\label{tab:baseline_results}
\resizebox{0.48\textwidth}{!}{
\begin{tabular}{l c c c}
\toprule
\textbf{Method} & \textbf{Halluc. $\downarrow$} & \textbf{Accuracy $\uparrow$} & \textbf{Quality $\uparrow$} \\
\midrule
Single-Agent & 0.417 & 0.792 & 0.763 \\
CoT & 0.398 & 0.801 & 0.772 \\
RAG & 0.352 & 0.815 & 0.781 \\
\midrule
\textbf{Multi-Agent (Ours)} & \textbf{0.272} & 0.769 & 0.737 \\
\bottomrule
\end{tabular}
}
\end{table}
Table~\ref{tab:baseline_results} shows that the proposed multi-agent cascade achieves the lowest hallucination score. Hallucination decreases from 0.417 in the Single-Agent baseline to 0.272 in the proposed method, indicating that sequential agent interaction provides more correction than isolated generation. CoT improves accuracy and quality relative to Single-Agent generation, but its hallucination score remains high at 0.398, suggesting that explicit reasoning improves structure without fully removing unsupported claims. RAG reduces hallucination to 0.352 and achieves the highest factual accuracy (0.815) and response quality (0.781), confirming the value of external grounding. However, the proposed cascade achieves hallucination reduction, suggesting that multi-agent refinement more directly impacts how hallucinated content is attenuated, preserved, and transformed across sequential interactions.
The results also reveal a clear trade-off. Although the proposed cascade reduces hallucinations, it has lower factual accuracy and response quality than RAG and CoT. This aligns with the chain-dynamics results: multi-agent refinement suppresses hallucination by constraining generated content, but may also remove valid factual detail and reduce semantic improvement. Therefore, the proposed method is for hallucination attenuation, while RAG remains for factual completeness and response quality. Additionally, hallucination mitigation is a multi-objective problem. Direct generation is efficient but more vulnerable to hallucination; CoT improves reasoning structure; RAG improves factual grounding; and the proposed cascade provides hallucination attenuation by modeling hallucination as a dynamic process across agents.

\section{Discussion}
\label{Discussion}
The results show that hallucination in multi-agent LLM systems exhibits structured propagation rather than an isolated generation error. Across the evaluated cascades, hallucination consistently decreases from earlier to later agents, indicating that sequential refinement can suppress unsupported, weakly grounded, and inconsistent claims. Later agents, therefore, act as corrective filters that revise, weaken, and remove unreliable content inherited from previous outputs. This attenuation, however, comes with a measurable cost. Lower hallucination is accompanied by declines in factual accuracy and response quality, showing that refinement not only removes errors, but it also reshapes the response. As agents revise earlier outputs, they may omit valid details, simplify nuanced claims, and replace specific content with safer but less informative statements. Thus, hallucination reduction must be interpreted in light of factual preservation, semantic richness, and response utility. Semantic drift is a key factor in this process. Higher drift is associated with higher hallucination, greater cascade risk, and lower quality. When a response moves away from the original prompt, later agents may treat the shifted content as valid context, allowing distorted meanings to persist across the cascade. In contrast, stronger information retention stabilizes the process by preserving factual continuity and limiting unnecessary rewriting. These findings suggest that reliable multi-agent behavior depends on balancing three forces: correction, drift control, and information retention. Weak correction allows hallucinations to persist, excessive correction can remove useful content, and uncontrolled drift can move the response away from the intended meaning. Effective cascade design should therefore reduce hallucination while preserving factual completeness, semantic stability, and useful response quality. Additionally, hallucination in multi-agent LLM systems should be analyzed as a dynamic system-level behavior. The reliability of a final answer depends not only on its final score, but also on how claims were introduced, transformed, corrected, and preserved across agents.

\subsection{Limitations}
\label{Limitations}
Despite the strengths of the proposed approach, several limitations should be acknowledged. First, the claim extraction process assumes that generated text can be accurately decomposed into atomic factual units. In practice, claim extraction may introduce errors and inconsistencies, especially in complex and ambiguous sentences. Second, the grounding mechanism relies on the availability and completeness of the reference knowledge base. In domains with incomplete and noisy knowledge, grounding accuracy may degrade, impacting hallucination estimation. Third, the model-based estimator depends on learned representations that may inherit biases from the underlying language model. This may lead to over- and under-estimation of hallucination in certain contexts. Fourth, the experimental evaluation focuses on a limited set of models and topics. While the results demonstrate consistent trends, broader validation across additional domains and model architectures is required. Furthermore, the current study does not explicitly account for adversarial inputs and prompt-injection scenarios, which may significantly alter the dynamics of hallucinations in real-world deployments.

\subsection{Future Work}
\label{Future Work}
Several directions can be explored to extend this work. First, future research can focus on integrating explicit control mechanisms to regulate hallucination propagation. For example, adaptive agent selection and confidence-aware routing strategies may help balance hallucination reduction with information retention. Second, extending the approach to adversarial settings is an important direction. Studying how prompt injection and malicious inputs impact cascade dynamics can improve the robustness of multi-agent systems. Third, incorporating energy and computational cost into the analysis would enable a joint evaluation of hallucination, performance, and sustainability, aligning with emerging Green AI objectives. Fourth, learning-based estimation of propagation parameters (e.g., autoregressive coefficients) can provide deeper insights into system stability and enable predictive modeling of hallucination trajectories. Moreover, applying the proposed approach to real-world applications such as healthcare, cybersecurity, and scientific reasoning can validate its practical utility and reveal domain-specific dynamics.

\section{Conclusion}
\label{Conclusion}
This paper presented a method for modeling hallucination in multi-agent LLM systems as a stochastic dynamical process. By decomposing generated text into atomic claims and combining rule-based grounding with model-based estimation, the proposed method enables fine-grained quantification of hallucinations. We further modeled multi-agent interaction as a cascade and introduced propagation metrics to analyze how hallucinations evolve across agents in sequence. The experimental results show that hallucination consistently decreases across cascade steps, indicating that multi-agent interaction can support structured correction. However, this reduction is accompanied by measurable declines in factual accuracy and response quality, revealing a trade-off between error suppression and information preservation. These findings show that hallucination is not only a static property of individual model outputs, but a dynamic signal that propagates, attenuates, and transforms through agent interactions. This perspective supports the design of more reliable, controllable, and interpretable multi-agent LLM systems.

\bibliographystyle{IEEEtran}
\bibliography{Ref}

@inproceedings{truthfulqa,
  title={TruthfulQA: Measuring How Models Mimic Human Falsehoods},
  author={Lin, Stephanie and Hilton, Jacob and Evans, Owain},
  booktitle={Advances in Neural Information Processing Systems (NeurIPS)},
  year={2022}
}

@article{selfcheckgpt,
  title={SelfCheckGPT: Zero-Resource Black-Box Hallucination Detection for Generative Large Language Models},
  author={Manakul, Potsawee and Liusie, Adian and Gales, Mark J. F.},
  journal={arXiv preprint arXiv:2303.08896},
  year={2023}
}

@article{survey_hallu,
  title={Survey of Hallucination in Natural Language Generation},
  author={Ji, Ziwei and Lee, Nayeon and Frieske, Rita and Yu, Tiezheng and Su, Dan and Xu, Yan and Ishii, Etsuko and Bang, Yejin and Madotto, Andrea and Fung, Pascale},
  journal={ACM Computing Surveys},
  volume={55},
  number={12},
  pages={1--38},
  year={2023},
  publisher={ACM}
}

@inproceedings{geval,
  title={G-Eval: NLG Evaluation using GPT-4 with Better Human Alignment},
  author={Liu, Yang and Iter, Dan and Xu, Yichong and Wang, Shuohang and Xu, Ruochen and Zhu, Chenguang},
  booktitle={Proceedings of the 2023 Conference on Empirical Methods in Natural Language Processing (EMNLP)},
  year={2023}
}

@article{medical_hallu,
  title={Hallucinations in Medical Devices},
  author={Granstedt, Jason and Kc, Prabhat and Deshpande, Rucha and Garcia, Victor and Badano, Aldo},
  journal={Artificial Intelligence in the Life Sciences},
  volume={8},
  pages={100145},
  year={2025},
  publisher={Elsevier}
}

@article{kg_llm,
  title={Knowledge Graphs, Large Language Models, and Hallucinations: An NLP Perspective},
  author={Lavrinovics, Ernests and Biswas, Russa and Bjerva, Johannes and Hose, Katja},
  journal={Web Semantics: Science, Services and Agents on the World Wide Web},
  volume={85},
  pages={100844},
  year={2025},
  publisher={Elsevier}
}

@inproceedings{kg_sysml,
  title={Mitigating Hallucinations in SysML v2 Generation Using LLMs and a Tri-Layered Knowledge Graph Reasoning Framework},
  author={Qualis, Richard A.},
  booktitle={2025 ACM/IEEE International Conference on Model Driven Engineering Languages and Systems Companion (MODELS-C)},
  year={2025},
  organization={IEEE}
}

@article{zhang2024knowhalu,
  title={Knowhalu: Hallucination detection via multi-form knowledge based factual checking},
  author={Zhang, Jiawei and Xu, Chejian and Gai, Yu and Lecue, Freddy and Song, Dawn and Li, Bo},
  journal={arXiv preprint arXiv:2404.02935},
  year={2024}
}

@article{yang2502hallucination,
  title={Hallucination detection in large language models with metamorphic relations, 2025a},
  author={Yang, Borui and Mamun, MAA and Zhang, Jie M and Uddin, Gias},
  journal={URL https://arxiv. org/abs/2502.15844}
}

@article{paudel2025hallucinot,
  title={Hallucinot: Hallucination detection through context and common knowledge verification},
  author={Paudel, Bibek and Lyzhov, Alexander and Joshi, Preetam and Anand, Puneet},
  journal={arXiv preprint arXiv:2504.07069},
  year={2025}
}

@article{farquhar2024detecting,
  title={Detecting hallucinations in large language models using semantic entropy},
  author={Farquhar, Sebastian and Kossen, Jannik and Kuhn, Lorenz and Gal, Yarin},
  journal={Nature},
  volume={630},
  number={8017},
  pages={625--630},
  year={2024},
  publisher={Nature Publishing Group UK London}
}

@inproceedings{binkowski2025hallucination,
  title={Hallucination detection in llms using spectral features of attention maps},
  author={Binkowski, Jakub and Janiak, Denis and Sawczyn, Albert and Gabrys, Bogdan and Kajdanowicz, Tomasz Jan},
  booktitle={Proceedings of the 2025 Conference on Empirical Methods in Natural Language Processing},
  pages={24365--24396},
  year={2025}
}

@inproceedings{karanikolas2023large,
  title={Large language models versus natural language understanding and generation},
  author={Karanikolas, Nikitas and Manga, Eirini and Samaridi, Nikoletta and Tousidou, Eleni and Vassilakopoulos, Michael},
  booktitle={Proceedings of the 27th Pan-Hellenic Conference on Progress in Computing and Informatics},
  pages={278--290},
  year={2023}
}

@article{ren2024advancements,
  title={Advancements and applications of large language models in natural language processing: A comprehensive review},
  author={Ren, Mengchao},
  journal={Applied and Computational Engineering},
  volume={97},
  pages={55--63},
  year={2024}
}

@article{hadi2023large,
  title={Large language models: a comprehensive survey of its applications, challenges, limitations, and future prospects},
  author={Hadi, Muhammad Usman and Qureshi, Rizwan and Shah, Abbas and Irfan, Muhammad and Zafar, Anas and Shaikh, Muhammad Bilal and Akhtar, Naveed and Wu, Jia and Mirjalili, Seyedali and others},
  journal={Authorea preprints},
  volume={1},
  number={3},
  pages={1--26},
  year={2023}
}

@inproceedings{gu2023don,
  title={Don’t generate, discriminate: A proposal for grounding language models to real-world environments},
  author={Gu, Yu and Deng, Xiang and Su, Yu},
  booktitle={Proceedings of the 61st annual meeting of the association for computational linguistics (volume 1: long papers)},
  pages={4928--4949},
  year={2023}
}

@article{augenstein2024factuality,
  title={Factuality challenges in the era of large language models and opportunities for fact-checking},
  author={Augenstein, Isabelle and Baldwin, Timothy and Cha, Meeyoung and Chakraborty, Tanmoy and Ciampaglia, Giovanni Luca and Corney, David and DiResta, Renee and Ferrara, Emilio and Hale, Scott and Halevy, Alon and others},
  journal={Nature Machine Intelligence},
  volume={6},
  number={8},
  pages={852--863},
  year={2024},
  publisher={Nature Publishing Group UK London}
}

@article{mcintosh2025inadequacies,
  title={Inadequacies of large language model benchmarks in the era of generative artificial intelligence},
  author={McIntosh, Timothy R and Susnjak, Teo and Arachchilage, Nalin and Liu, Tong and Xu, Dan and Watters, Paul and Halgamuge, Malka N},
  journal={IEEE Transactions on Artificial Intelligence},
  year={2025},
  publisher={IEEE}
}

@article{gao2025h,
  title={H-Neurons: On the Existence, Impact, and Origin of Hallucination-Associated Neurons in LLMs},
  author={Gao, Cheng and Chen, Huimin and Xiao, Chaojun and Chen, Zhiyi and Liu, Zhiyuan and Sun, Maosong},
  journal={arXiv preprint arXiv:2512.01797},
  year={2025}
}

@article{kulkarni2025scientific,
  title={Scientific hypothesis generation and validation: Methods, datasets, and future directions},
  author={Kulkarni, Adithya and Alotaibi, Fatimah and Zeng, Xinyue and Wu, Longfeng and Zeng, Tong and Yao, Barry Menglong and Liu, Minqian and Zhang, Shuaicheng and Huang, Lifu and Zhou, Dawei},
  journal={arXiv preprint arXiv:2505.04651},
  year={2025}
}

@article{huang2025survey,
  title={A survey on hallucination in large language models: Principles, taxonomy, challenges, and open questions},
  author={Huang, Lei and Yu, Weijiang and Ma, Weitao and Zhong, Weihong and Feng, Zhangyin and Wang, Haotian and Chen, Qianglong and Peng, Weihua and Feng, Xiaocheng and Qin, Bing and others},
  journal={ACM Transactions on Information Systems},
  volume={43},
  number={2},
  pages={1--55},
  year={2025},
  publisher={ACM New York, NY}
}

@article{tonmoy2024comprehensive,
  title={A comprehensive survey of hallucination mitigation techniques in large language models},
  author={Tonmoy, SM and Zaman, SM and Jain, Vinija and Rani, Anku and Rawte, Vipula and Chadha, Aman and Das, Amitava},
  journal={arXiv preprint arXiv:2401.01313},
  year={2024}
}

@article{lu2026loki,
  title={Loki’s dance of illusions: A comprehensive survey of hallucination in large language models},
  author={Lu, Ming and Li, Chaozhuo and Wang, Pengbo and Wang, Chenxu and Zhang, Litian and Liu, Zheng and Ye, Qiwei and Hua, Yi and Cai, Yushan and Xu, Yuanbo and others},
  journal={IEEE Transactions on Computational Social Systems},
  year={2026},
  publisher={IEEE}
}

@article{mckight2010kruskal,
  title={Kruskal-wallis test},
  author={McKight, Patrick E and Najab, Julius},
  journal={The corsini encyclopedia of psychology},
  pages={1--1},
  year={2010},
  publisher={Wiley Online Library}
}

@article{woolson2007wilcoxon,
  title={Wilcoxon signed-rank test},
  author={Woolson, Robert F},
  journal={Wiley encyclopedia of clinical trials},
  pages={1--3},
  year={2007},
  publisher={Wiley Online Library}
}
\end{document}